\documentclass[11pt]{article}

\newcommand{\bra}[1]{\langle #1|}
\newcommand{\ket}[1]{|#1\rangle}

\usepackage[margin=2.8cm, a4paper]{geometry}
\usepackage{enumitem}
\usepackage{dsfont}
\usepackage{amsmath,amsthm,amsfonts,amssymb}
\usepackage{fontenc}
\usepackage[all]{xy}
\usepackage{graphicx,subfigure,multirow}
\usepackage[figurename=Fig.]{caption}
\usepackage{tikz}
\usetikzlibrary{shapes, arrows, shadows}
\usepackage[latin1]{inputenc}
\usepackage[scaled]{helvet}
\usepackage[toc, page]{appendix}
\usepackage{dsfont}
\newtheorem{thm}{Theorem}[subsection]
\newtheorem{define}[thm]{Definition}

\newtheorem{lem}[thm]{Lemma}

\numberwithin{equation}{subsection}
\usepackage{cancel}

\usepackage[colorinlistoftodos]{todonotes}

\usepackage{tikz}
\pgfdeclarelayer{nodelayer}
\pgfdeclarelayer{edgelayer}
\pgfsetlayers{edgelayer,nodelayer,main}
\tikzstyle{none}=[inner sep=0pt]
\tikzstyle{new}=[inner sep=2pt]
\usetikzlibrary{shapes, arrows, shadows}
\usetikzlibrary{circuits.ee.IEC}
\usetikzlibrary{decorations.pathmorphing}
\usetikzlibrary{shapes.geometric}

\tikzstyle{env}=[copoint,regular polygon rotate=0,minimum width=0.2cm, fill=black]

\tikzstyle{probs}=[shape=semicircle,fill=white,draw=black,shape border rotate=180,minimum width=1.2cm]

\usetikzlibrary{circuits.ee.IEC}
\usetikzlibrary{decorations.pathmorphing}
\usetikzlibrary{shapes.geometric}

\tikzstyle{wavy}=[decorate,decoration={snake, segment length=1mm, amplitude=0.3mm}]
\tikzstyle{mopoint}=[shape=semicircle, fill=white,draw=black,shape border rotate=180,scale =0.75]
\tikzstyle{mocopoint}=[shape=semicircle, fill=white,draw=black,minimum width = 0.9cm, scale =0.75, xscale=0.7]
\tikzstyle{cpoint}=[shape=semicircle, fill=white,draw=black,minimum width = 0.9cm, scale =0.75, xscale=1, yscale=0.7, shape border rotate = 90,font=\fontsize{14}{16}\selectfont]
\tikzstyle{cocpoint}=[shape=semicircle, fill=white,draw=black,minimum width = 0.9cm, scale =0.75, xscale=1, yscale=0.7, shape border rotate = 270,font=\fontsize{14}{16}\selectfont]

\tikzstyle{every picture}=[baseline=-0.25em,scale=0.5]
\tikzstyle{dotpic}=[] 
\tikzstyle{diredges}=[every to/.style={diredge}]
\tikzstyle{math matrix}=[matrix of math nodes,left delimiter=(,right delimiter=),inner sep=2pt,column sep=1em,row sep=0.5em,nodes={inner sep=0pt},text height=1.5ex, text depth=0.25ex]

\tikzstyle{inline text}=[text height=1.5ex, text depth=0.25ex,yshift=0.5mm]
\tikzstyle{label}=[font=\footnotesize,text height=1.5ex, text depth=0.25ex,yshift=0.5mm]
\tikzstyle{left label}=[label,anchor=east,xshift=1.5mm]
\tikzstyle{right label}=[label,anchor=west,xshift=-1.5mm]

\tikzstyle{braceedge}=[decorate,decoration={brace,amplitude=2mm,raise=-1mm}]
\tikzstyle{small braceedge}=[decorate,decoration={brace,amplitude=1mm,raise=-1mm}]

\tikzstyle{doubled}=[line width=2pt] 
\tikzstyle{boldedge}=[doubled,shorten <=-0.17mm,shorten >=-0.17mm]

\tikzstyle{boldedgedashed}=[very thick,dashed,shorten <=-0.17mm,shorten >=-0.17mm]
\tikzstyle{vboldedgedashed}=[doubled,dashed,shorten <=-0.17mm,shorten >=-0.17mm]
\tikzstyle{left hook arrow}=[left hook-latex]
\tikzstyle{right hook arrow}=[right hook-latex]
\tikzstyle{sembracket}=[line width=0.5pt,shorten <=-0.07mm,shorten >=-0.07mm]

\tikzstyle{causal edge}=[->,thick,gray]
\tikzstyle{causal nondir}=[thick,gray]
\tikzstyle{timeline}=[thick,gray, dashed]

\tikzstyle{cedge}=[<->,thick,gray!70!white]

\tikzstyle{empty diagram}=[draw=gray!40!white,dashed,shape=rectangle,minimum width=1cm,minimum height=1cm]
\tikzstyle{empty diagram small}=[draw=gray!50!white,dashed,shape=rectangle,minimum width=0.6cm,minimum height=0.5cm]

\tikzstyle{dot}=[inner sep=0.7mm,minimum width=0pt,minimum height=0pt,draw,shape=circle]
\tikzstyle{ddot}=[inner sep=0.7mm,doubled, minimum width=2.5mm,minimum height=2.5mm,draw,shape=circle]

\tikzstyle{black dot}=[dot,fill=black]
\tikzstyle{white dot}=[dot,fill=white]
\tikzstyle{green dot}=[white dot] 
\tikzstyle{gray dot}=[dot,fill=gray!40!white]
\tikzstyle{red dot}=[gray dot] 

\tikzstyle{black ddot}=[ddot,fill=black]
\tikzstyle{white ddot}=[ddot,fill=white]
\tikzstyle{gray ddot}=[ddot,fill=gray!40!white]

\tikzstyle{gray edge}=[gray!40!white]

\tikzstyle{small dot}=[inner sep=0.4mm,minimum width=0pt,minimum height=0pt,draw,shape=circle]

\tikzstyle{small black dot}=[small dot,fill=black]
\tikzstyle{small white dot}=[small dot,fill=white]
\tikzstyle{small gray dot}=[small dot,fill=gray!40!white]

\tikzstyle{causal dot}=[inner sep=0.4mm,minimum width=0pt,minimum height=0pt,draw=white,shape=circle,fill=gray!40!white]

\tikzstyle{white phase dot}=[dot,fill=white]
\tikzstyle{white phase ddot}=[ddot,fill=white]

\tikzstyle{gray phase dot}=[dot,fill=gray!40!white]
\tikzstyle{gray phase ddot}=[ddot,fill=gray!40!white]
\tikzstyle{grey phase dot}=[gray phase dot]
\tikzstyle{grey phase ddot}=[gray phase ddot]

\tikzstyle{cnot}=[fill=white,shape=circle,inner sep=-1.4pt]
\tikzstyle{hadamard}=[square box,inner sep=0 pt,font=\tiny\sf,minimum height=3mm,minimum width=3mm]
\tikzstyle{dhadamard}=[hadamard,doubled]
\tikzstyle{antipode}=[white dot,inner sep=0.3mm,font=\footnotesize]

\tikzstyle{scalar}=[diamond,draw,inner sep=0.5pt,font=\small]
\tikzstyle{dscalar}=[diamond,doubled, draw,inner sep=0.5pt,font=\small]

\tikzstyle{small box}=[rectangle,inline text,fill=white,draw,minimum height=5mm,yshift=-0.5mm,minimum width=5mm,font=\small]
\tikzstyle{small gray box}=[small box,fill=gray!30]
\tikzstyle{medium box}=[rectangle,inline text,fill=white,draw,minimum height=5mm,yshift=-0.5mm,minimum width=10mm,font=\small]
\tikzstyle{square box}=[small box] 
\tikzstyle{medium gray box}=[small box,fill=gray!30]
\tikzstyle{large box}=[rectangle,inline text,fill=white,draw,minimum height=5mm,yshift=-0.5mm,minimum width=15mm,font=\small]
\tikzstyle{large gray box}=[small box,fill=gray!30]

\tikzstyle{point}=[regular polygon,regular polygon sides=3,draw,scale=0.75,inner sep=-0.5pt,minimum width=9mm,fill=white,regular polygon rotate=180]
\tikzstyle{copoint}=[regular polygon,regular polygon sides=3,draw,scale=0.75,inner sep=-0.5pt,minimum width=9mm,fill=white]
\tikzstyle{dpoint}=[point,doubled]
\tikzstyle{dcopoint}=[copoint,doubled]

\tikzstyle{tinypoint}=[regular polygon,regular polygon sides=3,draw,scale=0.55,inner sep=-0.15pt,minimum width=6mm,fill=white,regular polygon rotate=180]

\tikzstyle{white point}=[point]
\tikzstyle{green point}=[white point] 
\tikzstyle{white copoint}=[copoint]
\tikzstyle{gray point}=[point,fill=gray!40!white]
\tikzstyle{gray dpoint}=[gray point,doubled]
\tikzstyle{red point}=[gray point] 
\tikzstyle{gray copoint}=[copoint,fill=gray!40!white]
\tikzstyle{gray dcopoint}=[gray copoint,doubled]

\tikzstyle{tiny gray point}=[tinypoint,fill=gray!40!white]

\tikzstyle{diredge}=[->]
\tikzstyle{rdiredge}=[<-]
\tikzstyle{thickdiredge}=[->, very thick]
\tikzstyle{pointer edge}=[->,very thick,gray]
\tikzstyle{pointer edge part}=[very thick,gray]
\tikzstyle{dashed edge}=[dashed]
\tikzstyle{thick dashed edge}=[very thick,dashed]
\tikzstyle{thick gray dashed edge}=[thick dashed edge,gray!90]
\tikzstyle{thick map edge}=[very thick,|->]

\makeatletter
\newcommand{\boxshape}[3]{%
\pgfdeclareshape{#1}{
\inheritsavedanchors[from=rectangle] 
\inheritanchorborder[from=rectangle]
\inheritanchor[from=rectangle]{center}
\inheritanchor[from=rectangle]{north}
\inheritanchor[from=rectangle]{south}
\inheritanchor[from=rectangle]{west}
\inheritanchor[from=rectangle]{east}
\backgroundpath{
\southwest \pgf@xa=\pgf@x \pgf@ya=\pgf@y
\northeast \pgf@xb=\pgf@x \pgf@yb=\pgf@y

\@tempdima=#2
\@tempdimb=#3

\pgfpathmoveto{\pgfpoint{\pgf@xa - 5pt + \@tempdima}{\pgf@ya}}
\pgfpathlineto{\pgfpoint{\pgf@xa - 5pt - \@tempdima}{\pgf@yb}}
\pgfpathlineto{\pgfpoint{\pgf@xb + 5pt + \@tempdimb}{\pgf@yb}}
\pgfpathlineto{\pgfpoint{\pgf@xb + 5pt - \@tempdimb}{\pgf@ya}}
\pgfpathlineto{\pgfpoint{\pgf@xa - 5pt + \@tempdima}{\pgf@ya}}
\pgfpathclose
}
}}

\boxshape{NEbox}{0pt}{5pt}
\boxshape{SEbox}{0pt}{-5pt}
\boxshape{NWbox}{5pt}{0pt}
\boxshape{SWbox}{-5pt}{0pt}
\boxshape{EBox}{-3pt}{3pt}
\boxshape{WBox}{3pt}{-3pt}
\makeatother

\tikzstyle{cloud}=[shape=cloud,draw,minimum width=1.5cm,minimum height=1.5cm]

\tikzstyle{map}=[draw,shape=NEbox,inner sep=2pt,minimum height=6mm,fill=white]
\tikzstyle{mapdag}=[draw,shape=SEbox,inner sep=2pt,minimum height=6mm,fill=white]
\tikzstyle{mapadj}=[draw,shape=SEbox,inner sep=2pt,minimum height=6mm,fill=white]
\tikzstyle{maptrans}=[draw,shape=SWbox,inner sep=2pt,minimum height=6mm,fill=white]
\tikzstyle{mapconj}=[draw,shape=NWbox,inner sep=2pt,minimum height=6mm,fill=white]

\tikzstyle{dbox}=[draw,doubled,shape=rectangle,inner sep=2pt,minimum height=6mm,minimum width=6mm,fill=white]
\tikzstyle{dmap}=[draw,doubled,shape=NEbox,inner sep=2pt,minimum height=6mm,fill=white]
\tikzstyle{dmapdag}=[draw,doubled,shape=SEbox,inner sep=2pt,minimum height=6mm,fill=white]
\tikzstyle{dmapadj}=[draw,doubled,shape=SEbox,inner sep=2pt,minimum height=6mm,fill=white]
\tikzstyle{dmaptrans}=[draw,doubled,shape=SWbox,inner sep=2pt,minimum height=6mm,fill=white]
\tikzstyle{dmapconj}=[draw,doubled,shape=NWbox,inner sep=2pt,minimum height=6mm,fill=white]

\tikzstyle{ddmap}=[draw,doubled,dashed,shape=NEbox,inner sep=2pt,minimum height=6mm,fill=white]
\tikzstyle{ddmapdag}=[draw,doubled,dashed,shape=SEbox,inner sep=2pt,minimum height=6mm,fill=white]
\tikzstyle{ddmapadj}=[draw,doubled,dashed,shape=SEbox,inner sep=2pt,minimum height=6mm,fill=white]
\tikzstyle{ddmaptrans}=[draw,doubled,dashed,shape=SWbox,inner sep=2pt,minimum height=6mm,fill=white]
\tikzstyle{ddmapconj}=[draw,doubled,dashed,shape=NWbox,inner sep=2pt,minimum height=6mm,fill=white]

\boxshape{sNEbox}{0pt}{3pt}
\boxshape{sSEbox}{0pt}{-3pt}
\boxshape{sNWbox}{3pt}{0pt}
\boxshape{sSWbox}{-3pt}{0pt}
\tikzstyle{smap}=[draw,shape=sNEbox,fill=white]
\tikzstyle{smapdag}=[draw,shape=sSEbox,fill=white]
\tikzstyle{smapadj}=[draw,shape=sSEbox,fill=white]
\tikzstyle{smaptrans}=[draw,shape=sSWbox,fill=white]
\tikzstyle{smapconj}=[draw,shape=sNWbox,fill=white]

\tikzstyle{dsmap}=[draw,dashed,shape=sNEbox,fill=white]
\tikzstyle{dsmapdag}=[draw,dashed,shape=sSEbox,fill=white]
\tikzstyle{dsmaptrans}=[draw,dashed,shape=sSWbox,fill=white]
\tikzstyle{dsmapconj}=[draw,dashed,shape=sNWbox,fill=white]

\boxshape{mNEbox}{0pt}{10pt}
\boxshape{mSEbox}{0pt}{-10pt}
\boxshape{mNWbox}{10pt}{0pt}
\boxshape{mSWbox}{-10pt}{0pt}
\tikzstyle{mmap}=[draw,shape=mNEbox]
\tikzstyle{mmapdag}=[draw,shape=mSEbox]
\tikzstyle{mmaptrans}=[draw,shape=mSWbox]
\tikzstyle{mmapconj}=[draw,shape=mNWbox]

\tikzstyle{mmapgray}=[draw,fill=gray!40!white,shape=mNEbox]
\tikzstyle{smapgray}=[draw,fill=gray!40!white,shape=sNEbox]

\makeatletter
\pgfdeclareshape{cornerpoint}{
\inheritsavedanchors[from=rectangle] 
\inheritanchorborder[from=rectangle]
\inheritanchor[from=rectangle]{center}
\inheritanchor[from=rectangle]{north}
\inheritanchor[from=rectangle]{south}
\inheritanchor[from=rectangle]{west}
\inheritanchor[from=rectangle]{east}
\backgroundpath{
\southwest \pgf@xa=\pgf@x \pgf@ya=\pgf@y
\northeast \pgf@xb=\pgf@x \pgf@yb=\pgf@y

\pgfmathsetmacro{\pgf@shorten@left}{\pgfkeysvalueof{/tikz/shorten left}}
\pgfmathsetmacro{\pgf@shorten@right}{\pgfkeysvalueof{/tikz/shorten right}}

\pgfpathmoveto{\pgfpoint{0.5 * (\pgf@xa + \pgf@xb)}{\pgf@ya - 5pt}}
\pgfpathlineto{\pgfpoint{\pgf@xa - 8pt + \pgf@shorten@left}{\pgf@yb - 1.5 * \pgf@shorten@left}}
\pgfpathlineto{\pgfpoint{\pgf@xa - 8pt + \pgf@shorten@left}{\pgf@yb}}
\pgfpathlineto{\pgfpoint{\pgf@xb + 8pt - \pgf@shorten@right}{\pgf@yb}}
\pgfpathlineto{\pgfpoint{\pgf@xb + 8pt - \pgf@shorten@right}{\pgf@yb - 1.5 * \pgf@shorten@right}}
\pgfpathclose
}
}

\pgfdeclareshape{cornercopoint}{
\inheritsavedanchors[from=rectangle] 
\inheritanchorborder[from=rectangle]
\inheritanchor[from=rectangle]{center}
\inheritanchor[from=rectangle]{north}
\inheritanchor[from=rectangle]{south}
\inheritanchor[from=rectangle]{west}
\inheritanchor[from=rectangle]{east}
\backgroundpath{
\southwest \pgf@xa=\pgf@x \pgf@ya=\pgf@y
\northeast \pgf@xb=\pgf@x \pgf@yb=\pgf@y

\pgfmathsetmacro{\pgf@shorten@left}{\pgfkeysvalueof{/tikz/shorten left}}
\pgfmathsetmacro{\pgf@shorten@right}{\pgfkeysvalueof{/tikz/shorten right}}

\pgfpathmoveto{\pgfpoint{0.5 * (\pgf@xa + \pgf@xb)}{\pgf@yb + 5pt}}
\pgfpathlineto{\pgfpoint{\pgf@xa - 8pt + \pgf@shorten@left}{\pgf@ya + 1.5 * \pgf@shorten@left}}
\pgfpathlineto{\pgfpoint{\pgf@xa - 8pt + \pgf@shorten@left}{\pgf@ya}}
\pgfpathlineto{\pgfpoint{\pgf@xb + 8pt - \pgf@shorten@right}{\pgf@ya}}
\pgfpathlineto{\pgfpoint{\pgf@xb + 8pt - \pgf@shorten@right}{\pgf@ya + 1.5 * \pgf@shorten@right}}
\pgfpathclose
}
}

\makeatother

\pgfkeyssetvalue{/tikz/shorten left}{0pt}
\pgfkeyssetvalue{/tikz/shorten right}{0pt}

\tikzstyle{kpoint common}=[draw,fill=white,inner sep=1pt,minimum height=4mm]
\tikzstyle{kpoint}=[shape=cornerpoint,shorten left=5pt,kpoint common]
\tikzstyle{kpoint adjoint}=[shape=cornercopoint,shorten left=5pt,kpoint common]
\tikzstyle{kpoint conjugate}=[shape=cornerpoint,shorten right=5pt,kpoint common]
\tikzstyle{kpoint transpose}=[shape=cornercopoint,shorten right=5pt,kpoint common]
\tikzstyle{kpoint symm}=[shape=cornerpoint,shorten left=5pt,shorten right=5pt,kpoint common]

\tikzstyle{kpointdag}=[kpoint adjoint]
\tikzstyle{kpointadj}=[kpoint adjoint]
\tikzstyle{kpointconj}=[kpoint conjugate]
\tikzstyle{kpointtrans}=[kpoint transpose]

\tikzstyle{dkpoint}=[kpoint,doubled]
\tikzstyle{dkpointdag}=[kpoint adjoint,doubled]
\tikzstyle{dkcopoint}=[kpoint adjoint,doubled]
\tikzstyle{dkpointadj}=[kpoint adjoint,doubled]
\tikzstyle{dkpointconj}=[kpoint conjugate,doubled]
\tikzstyle{dkpointtrans}=[kpoint transpose,doubled]

\tikzstyle{kscalar}=[kpoint common, shape=EBox, inner xsep=-1pt, inner ysep=3pt,font=\small]
\tikzstyle{kscalarconj}=[kpoint common, shape=WBox, inner xsep=-1pt, inner ysep=3pt,font=\small]

 \tikzstyle{upground}=[circuit ee IEC,thick,ground,rotate=90,scale=2.5]
 \tikzstyle{downground}=[circuit ee IEC,thick,ground,rotate=-90,scale=2.5]
 \tikzstyle{bigground}=[regular polygon,regular polygon sides=3,draw=gray,scale=0.50,inner sep=-0.5pt,minimum width=10mm,fill=gray]

\tikzstyle{arrs}=[-latex,font=\small,auto]
\tikzstyle{arrow plain}=[arrs]
\tikzstyle{arrow dashed}=[dashed,arrs]
\tikzstyle{arrow bold}=[very thick,arrs]
\tikzstyle{arrow hide}=[draw=white!0,-]
\tikzstyle{arrow reverse}=[latex-]
\tikzstyle{cdnode}=[]

\tikzstyle{slit}=[line width=2]
\tikzstyle{block}=[line width=4,gray,line cap=round]
\tikzstyle{screen}=[line width=4,black,line cap=round]
\tikzstyle{di}=[diamond,draw,inner sep=0.5pt,font=\small, minimum size = .5cm]
\tikzstyle{sbox}=[rectangle,draw]
\tikzstyle{mirror}=[line width=2,black]
\tikzstyle{trace}=[circuit ee IEC,thick,ground,rotate=0,scale=2]
\tikzstyle{traceState}=[circuit ee IEC,thick,ground,rotate=180,scale=2]
\tikzstyle{detEff}=[circuit ee IEC,thick,ground,rotate=180,scale=1.4]
\tikzstyle{maxMix}=[circuit ee IEC,thick,ground,scale=1.4]
\tikzstyle{particlePath}=[line width=2,gray!40, line cap =round]
\usepackage{fp}
\usetikzlibrary{calc,decorations.pathmorphing,decorations.text,fixedpointarithmetic}

\newcommand{\detEff}{(\begin{tikzpicture}
	\begin{pgfonlayer}{nodelayer}
		\node [style=detEff] (0) at (0, -0) {};
		\node [style=none] (1) at (0.15, -0) {};
	\end{pgfonlayer}	
\end{tikzpicture}|}

\begin{document}

\begingroup
\centering
{\Large\textbf{Oracles and query lower bounds in generalised probabilistic theories} \\[1.5em]
 \normalsize Howard Barnum\textsuperscript{$\dagger,\mathsection$,}
\footnote{Electronic address: hnbarnum@aol.com}, Ciar\'{a}n
M. Lee\textsuperscript{$\ddagger$,}\footnote{ Electronic address:
  ciaran.lee@ucl.ac.uk}, John
H. Selby\textsuperscript{$\ast,\bullet$,}\footnote{Electronic address:
  john.selby08@imperial.ac.uk} } \\[1em]
\it 
Mathematical Sciences, University of Copenhagen,
Denmark.\\ 
\textsuperscript{$\mathsection$} Department of Physics and
Astronomy, University of New Mexico, Albuquerque,
USA. \\ 
\textsuperscript{$\ddagger$} Department of Physics and
Astronomy, University College London, UK. \\ 
\textsuperscript{$\ast$} University of Oxford, Department of Computer
Science, OX1 3QD, UK. \\ 
\textsuperscript{$\bullet$} Imperial College
London, London SW7 2AZ, UK.  \endgroup

\begin{abstract}
\small{
We investigate the connection between interference and computational
power within the operationally defined framework of generalised
probabilistic theories. To compare the computational abilities of
different theories within this framework we show that any theory
satisfying four natural physical principles possess a well-defined
oracle model. Indeed, we prove a subroutine theorem for oracles in
such theories which is a necessary condition for the oracle model to
be well-defined.  The four principles are: causality (roughly, no
signalling from the future), purification (each mixed state arises as
the marginal of a pure state of a larger system), strong symmetry
(existence of a rich set of nontrivial reversible transformations),
and informationally consistent composition (roughly: the information
capacity of a composite system is the sum of the capacities of its
constituent subsystems). Sorkin has defined a hierarchy of conceivable
interference behaviours, where the order in the hierarchy corresponds
to the number of paths that have an irreducible interaction in a
multi-slit experiment. Given our oracle model, we show that if a
classical computer requires at least $n$ queries to solve a learning
problem, because fewer queries provide no information about the
solution, then the corresponding ``no-information'' lower bound in
theories lying at the $k$th level of Sorkin's hierarchy is
$\lceil{n/k}\rceil$.  This lower bound leaves open the possibility
that quantum oracles are less powerful than general probabilistic
oracles, although it is not known whether the lower bound is
achievable in general.  Hence searches for higher-order interference
are not only foundationally motivated, but constitute a search for a
computational resource that might have power beyond that offered by
quantum computation.}

\end{abstract}

Landauer's Principle \cite{landauer1961irreversibility} states that
any logically irreversible processing or manipulation of information,
such as the erasure of a bit, must always be accompanied by an entropy
increase in the environment of the system processing the information.
As this illustrates, information is intimately tied to the physical
system that embodies it and is hence bound by physical
law---alternatively, \emph{information is physical}. If information
processing---or computation---is bound by physical law, then the
ultimate limits of computation should be derivable from natural
physical principles. Indeed, the advent of quantum computation
demonstrated that different physical principles lead to different
limits on computational power. This naturally leads to the question of
what general relationships hold between computational power and
physical principles. This question has recently been studied in the
framework of generalised probabilistic theories
\cite{paterek2010theories,lee2015computation,lee2015proofs,lee2016deriving,landscape,lee2016information},
which contains operationally-defined physical theories that generalise
the probabilistic formalism of quantum theory. By studying how
computational power varies as the underlying physical theory is
changed, one can determine the connection between physical principles
and computational power in a manner not tied to the specific
mathematical manifestation of a particular principle within a theory.

Most previous research into computation within the generalised
probabilistic theory framework has focused on deriving
general \emph{limitations} on computational ability from natural physical
principles\footnote{With the exception of \cite{landscape}, which
  constructs a theory capable of post-quantum computation. However,
  whether this computational advantage is directly tied to a simple
  physical principle remains unclear.}. No work to date has tied a
computational \emph{advantage} directly to a physical principle. For
instance, it is known that quantum interference between computational
paths is a resource for post-classical computation
\cite{stahlke2014quantum}, but it is not clear whether the presence of
interference in a general theory entails post-classical computation
\cite{lee2016generalised}, nor whether post-quantum interference
behaviour is in general a resource for post-quantum computation
\cite{lee2016deriving}. The former point concerns whether it is just
the particular mathematical description of interference in Hilbert
space which can be exploited to provide an advantage over classical
computation or whether such a statement can be seen to directly follow
from the observation
of interference in nature, and the latter concerns whether ``more''
interference implies, or at least can sometimes allow, ``more''
computational power.

Indeed, as first noted by Rafael Sorkin
\cite{sorkin1994quantum,sorkin1995quantum}, there is a limit to
quantum interference---at most \emph{pairs} of computational paths can ever
interact in a fundamental way. Sorkin has defined a hierarchy of
operationally conceivable interference behaviours---currently under
experimental investigation
\cite{sinha2008testing,park2012three,sinha2010ruling,kauten2015obtaining}---where
classical theory is at the first level of the hierarchy and quantum
theory belongs to the second. Informally, the order in the hierarchy
corresponds to the number of paths that have an irreducible
interaction in a multi-slit experiment. Given this definition, one can
investigate the role interference plays in computation in a
theory-independent manner by asking whether theories at level $k$
possess a computational advantage over theories at level $k-1, k-2,
\dots$.

One usually demonstrates the existence of a quantum advantage over
classical computation using \emph{oracles}. Indeed, the Deutsch-Jozsa
problem \cite{nielsen2010quantum} is such a example: given a function
$f:\{0,1\}^n\rightarrow\{0,1\}$, one is asked to determine whether it
is constant (same output for all inputs) or balanced ($0$ on exactly
half of the inputs)---promised that it is one of these
cases. Performance is quantified by the number of queries to an
oracle, which implements $f$ on any desired input, needed to solve the
problem. A deterministic classical computer requires $2^{n-1}+1$
queries, but Deutsch and Jozsa showed that a quantum computer can
solve the problem in a single query to an appropriately defined
\emph{quantum} oracle. Hence, to compare the computational abilities
of different theories within the framework a well-defined oracle model
is needed. We show that such a model can be defined in any theory
satisfying the following physical principles: \emph{causality}
(roughly, no signalling from the future), \emph{purification} (each
mixed state arises as the marginal of a pure state of a larger
system), and \emph{strong symmetry} (existence of a rich group of
non-trivial reversible transformations){\color{black}; additionally,
  we demand \emph{informationally consistent composition} (the act of
  bringing systems together cannot create or destroy
  information)}. Moreover, we prove a subroutine theorem for theories
satisfying these principles. That is, we show that having access to an
oracle for a particular decision problem which can be efficiently
solved in a given theory does not provide any more computational power
than just using the efficient algorithm itself. Such a result was
proved for quantum theory by Bennett et al. in
\cite{bennett1997strengths}, and is a necessary condition for a
well-defined oracle model.

Given this oracle model, we investigate whether lower bounds on the
number of queries needed by a quantum computer to solve certain
computational problems can be reduced in theories which possess
higher-order interference and satisfy the principles discussed
above. Indeed, we generalise results due to Meyer and Pommersheim
\cite{meyer2011uselessness} and show that if a particular type of
lower bound to a given query problem (from a fairly large class of
such problems) is $n$ using a classical computer, then the
corresponding lower bound for the same problem in theories with $k$th
order interference is $\lceil{n/k}\rceil$. As quantum theory only
exhibits second order interference, theories with post-quantum
interference allow for post-quantum computation. For example, in the
problem where we are asked to determine the parity of a function
$f:\{1,\dots,k\}\rightarrow\{0,1\}$ (which generalizes the special
case of the Deutsch-Jozsa problem for functions 
$f:\{0,1\} \rightarrow \{0,1\}$), $\lceil{k/2}\rceil$
quantum queries are needed, but we show that in any theory satisfying
our principles which has $k$th order interference, our generalisation
of the Meyer-Pommersheim useless-queries bound leaves open the
possibility that the parity can be determined with a single query.
This bound leaves open the possibility that the power of computation
might be improved by modifying interference behaviour in an
operationally conceivable manner. Hence searches for higher-order
interference are not only foundationally motivated, but constitute a
search for a computational resource potentially beyond that offered by
quantum computation.  {\color{black} An important direction for future
  work is to determine whether in theories satisfying these principles
  it is possible to find an algorithm that reaches this lower
  bound. We discuss potential ways to do this in the conclusion.  }

Other authors have considered computation beyond the usual quantum
formalism from a different perspective to the one employed here. For
example, Aaronson has considered alternate modifications of quantum
theory, such as a hidden variable model in which the history of hidden
states can be read out by the observer, and---together with
collaborators in \cite{aaronson2016space}---a model in which one is
given the ability to perform certain unphysical non-collapsing
measurements. Both of these models have been shown to entail
computational speed-ups over the usual quantum
formalism. Additionally, Bao et al. \cite{bao2015grover} have
investigated computation in modifications of quantum theory suggested
by the black hole information loss paradox and have shown the ability
to signal faster than light in such theories is intimately linked to a
speed-up over standard quantum theory in searching an unstructured
database. In contrast, the generalised probabilistic theory framework
employed here allowed for an investigation of query lower bounds and
computational advantages in alternate theories that are physically
reasonable and which, for instance, do not allow for superluminal
signalling \cite{barrett2007information}, cloning
\cite{barnum2007generalized}, or other phenomena that are arguably 
undesirable features of a theory.

The paper is organised as follows. In section~\ref{framework}, the
generalised probabilistic theory framework will be introduced along
with our physical principles and the definition of higher-order
interference. In section~\ref{section:oracles} the oracle model will
be introduced and defined and the subroutine theorem will be stated,
with the proof presented in appendix~\ref{Proof of
  subroutine}. Finally, in section~\ref{lower bound} we derive lower
bounds on query problems that directly follow from our principles.

\section{The framework} \label{framework}

One of the fundamental requirements of a physical theory is that it
should provide a consistent account of experimental observations. This
viewpoint underlies the framework of generalised probabilistic
theories
\cite{barrett2007information,chiribella2010probabilistic,hardy2011reformulating,lee2017no}. A
generalised probabilistic theory specifies a set of laboratory devices
that can be connected together in different ways to form experiments
and specifies probability distributions over experimental outcomes. A
device comes equipped with input ports, output ports, and a classical
pointer. When a device is used in an experiment, the classical pointer
comes to rest in one of a number of positions (``values''), indicating
an outcome has occurred. Intuitively, one envisages \emph{physical
  systems} passing between the ports of these devices. Such physical
systems come in different types, denoted $A,B,\dots$. One constructs
experiments by composing devices both sequentially and in parallel,
and when composed sequentially, types must match.

In this framework, closed circuits---those with no unconnected ports
and no cycles---are associated with probabilities---a probability for
each assigment of values to all classical pointers appearing in the
circuit.  These add to unity when summed over all assignments of
values to pointers in the circuit.  We use the term ``element'' for a
pair of device and pointer value; circuits may also be constructed of
such elements, and closed circuits of this type are associated with
individual probabilities.  The set of equivalence classes of elements
with no input ports are called \emph{states}, no output ports
\emph{effects} and both input and output ports \emph{transformations}.
The set of all states of system $A$ is denoted $\mathrm{St(A)}$, the
set of all effects on $B$ is denoted $\mathrm{Eff(B)}$ and the set of
transformations between systems $A$ and $B$ is denoted
$\mathrm{Transf(A,B)}$. Using standard operational assumptions and
arguments
\cite{lee2015computation,chiribella2010probabilistic,barrett2007information},
one can show that the set of states, effects and transformations each
give rise to a real vector space with transformations and effects
acting linearly on the real vector space of states. We assume in this
work that all vector spaces are finite
dimensional\footnote{Operationally this can be seen as saying that one
  does not need to perform an infinite number of distinct experiments
  to characterise states}.

A state is said to be \emph{pure} if it does not arise as a
\emph{coarse-graining} of other states\footnote{The process
  $\{\mathcal{U}_j\}_{j\in{Y}}$, where $j$ index the positions of the
  classical pointer, is a coarse-graining of the process
  $\{\mathcal{E}_i\}_{i\in{X}}$ if there is a disjoint partition
  $\{X_j\}_{j\in{Y}}$ of $X$ such that
  $\mathcal{U}_j=\sum_{i\in{X_j}}\mathcal{E}_i$.}; a pure state is one
for which we have maximal information. A state is \emph{mixed} if it
is not pure. A state $\omega$ is \emph{completely mixed} if for every pure
state $\chi$, $\omega$ can be expressed as a coarsegraining of a set
of states that includes $\chi$. Similarly, one says a transformation,
respectively an effect, is pure if it does not arise as a
coarse-graining of other transformations, respectively effects. It can
be shown that reversible transformations preserve pure states
\cite{chiribella2015entanglement}.  We'll say a measurement is pure if
all of its outcomes are pure effects. A state is \emph{maximally mixed} 
if, when expressed as a convex combination $\omega = \sum_{i \in S} p_i \omega_i$ of perfectly distinguishable
pure states, as is always possible given our assumptions, the probabilities
$p_i$ are uniform ($p_i = 1/|S|$).

The following `Dirac-like' notation $_A|s_i)$ will be used to
represent a state\footnote{or, more accurately, the real vector
  corresponding to the state.} of system $A$, and $(e_{r}|_B$ to
represent an effect on $B$. Here $i$ and $r$ represent the position of
the classical pointer associated to the device the prepares the state
and performs the measurement, respectively. The full measurement is
defined by the collection $\{(e_r|\}_r$. States, effects, and
transformations can be represented diagrammatically:
\[\begin{tikzpicture}
	\begin{pgfonlayer}{nodelayer}
		\node [style=cpoint] (0) at (-0.75, -0) {\huge{${s_i}$}};
		\node [style=small box,minimum size=0.75cm] (1) at (1.5, -0) {$T$};
		\node [style=cocpoint] (2) at (3.75, -0) {\huge{$e_r$}};
		\node [style=none] (3) at (5.5, -0) {$:=$};
		\node [style=none] (4) at (8.5, -0) {$(e_r|_BT_A|s)$};
		\node [style=none] (5) at (0.25, 0.5) {$A$};
		\node [style=none] (6) at (2.75, 0.5) {$B$};
		\node [style=none] (7) at (10.5, -0) {};
	\end{pgfonlayer}
	\begin{pgfonlayer}{edgelayer}
		\draw (0) to (1);
		\draw (1) to (2);
	\end{pgfonlayer}
\end{tikzpicture}\]
The above diagram represents the joint probability of preparing state
$|s_i)$, acting with transformation $T$ and registering outcome $r$
for the measurement $\{(e_r|\}_r$. In the above, the wires represent
physical systems, with their type denoted by the letter above
them. This diagrammatic representation was inspired by categorical
quantum mechanics \cite{coecke2016picturing,coecke2010quantum}.  Note 
that in the ``bra-ket'' like notation, the time-ordering (first
states, then transformations, then effects) ``flows'' from right to
left, while the reverse is true in the diagrammatic notation.

In the rest of the paper, it will be assumed that all theories satisfy the following physical principles.
\begin{define}[Deterministic effect, Causality \cite{chiribella2010probabilistic}]
An effect is called \emph{deterministic} if it has probability $1$ in
all states.  A theory is said to be \emph{causal} if there exists a
unique deterministic effect for every system.  We denote this effect
by $\detEff $.  In a causal theory, $\sum_r (e_r|=\detEff $ for all
measurements $\{(e_r|\}_r$ on a given system.
\end{define}
Mathematically, the principle of causality is equivalent to the
statement: ``Probabilities of outcomes of present experiments are
independent of future measurement choices''. In causal theories, all
states are \emph{normalised} \cite{chiribella2010probabilistic}. That
is, $\detEff s)=1$ for all $|s)$. Moreover, the unique deterministic
effect allows one to define a notion of \emph{marginalisation} for
multi-partite states.
\begin{define}[Purification \cite{chiribella2010probabilistic}] \label{Pure}
Given a state $_A|s)$ there exists a system $B$ and a pure state $_{AB}|\psi)$ on $AB$ such that $_A|s)$ is the marginalisation of $_{AB}|\psi)$: \[\begin{tikzpicture}
	\begin{pgfonlayer}{nodelayer}
		\node [style=none] (0) at (1, -0) {};
		\node [style=none] (1) at (1, 1.5) {};
		\node [style=none] (2) at (1, 2) {};
		\node [style=none] (3) at (1, -0.5) {};
		\node [style=trace] (4) at (2.5, -0) {};
		\node [style=none] (5) at (2.5, 1.5) {};
		\node [style=none] (6) at (1, -0.5) {};
		\node [style=none] (7) at (0.5, 0.75) {$\psi$};
		\node [style=none] (8) at (4, 1) {$=$};
		\node [style=cpoint] (9) at (5.5, 1.5) {$s$};
		\node [style=none] (10) at (7, 1.5) {};
		\node [style=none] (11) at (1.75, 0.5) {$B$};
		\node [style=none] (12) at (1.75, 2) {$A$};
		\node [style=none] (13) at (6.25, 2) {$A$};
		\node [style=none] (14) at (7.5, -0) {};
	\end{pgfonlayer}
	\begin{pgfonlayer}{edgelayer}
		\draw [bend right=90, looseness=1.25] (2.center) to (3.center);
		\draw (2.center) to (3.center);
		\draw (1.center) to (5.center);
		\draw (0.center) to (4);
		\draw (9) to (10.center);
	\end{pgfonlayer}
\end{tikzpicture} \] Moreover, the purification $_{AB}|\psi)$ is unique up to reversible transformations on the purifying system, $B$. That is if two states $|\psi)_{AB}$ and $|\psi')_{AB}$ purify $|s)_A$, then there exists a reversible transformation $T_B$ on system $B$ such that $_{AB}|\psi)=(\mathbb{I}_A\otimes{T_B})~_{AB}|\psi)$.
\end{define}
As pure states are those in which we have maximal information about a
system,\footnote{In the following sense: a state that is not pure can
  be written as a convex combination of other states, which can be
  thought of as a lack of information about which of these alternative
  states describes the system; of course in general in a nonclassical
  theory there are many, incompatible, such convex decompositions of
  the state.  But a pure state admits no such decomposition.}
purification principle formalises the statement that each state of
incomplete information about a system can arise in an essentially
unique way due to a lack of full access to a larger system that it is
part of.  In \cite{chiribella2015conservation} it was argued that
purification can be thought of as a statement of ``information
conservation''.  In a theory with purification, any missing
information about the state of a given system can always be traced
back to lack of access to some environment system.

We introduce one final principle which ensures that the information
stored in a system is compatible with composition, that is, we demand
that the mere act of bringing systems together should not create or
destroy information.\footnote{Some may prefer another way of glossing
  this principle: that the capacity of a pair of systems to store
  classical information should be the same whether they are accessed
  separately or jointly.}  If this were not the case then one could
potentially use this new global degree of freedom, representing the
increase of information capacity, to hide solutions to a hard
computational problem allowing one to solve a hard problem that could
not be solved by using the systems independently
\cite{lee2015computation}. We formalise this as follows:



\begin{define}[Informationally consistent composition] This consists of two constraints on parallel composition: i) the product of pure states is pure, ii) the product of maximally mixed states is maximally mixed.
\end{define}
The first of these formalises the intuitive idea that if one has
maximal information about each of two systems, then one has maximal
information about the composite of the two systems.
The existence of a maximally
mixed state, that is, a state about which we have minimal information,
is guaranteed for each system by purification
\cite{chiribella2010probabilistic}.  



The purification principle, in conjunction with causality and the
constraints on parallel composition discussed above, implies many
quantum information processing \cite{chiribella2010probabilistic} and
computational primitives \cite{lee2016deriving}. Examples include
teleportation, no information without disturbance, and no-bit
commitment \cite{chiribella2010probabilistic}. Moreover, purification
also leads to a well-defined notion of thermodynamics
\cite{chiribella2015entanglement,chiribella2016entanglement,chiribella2016purity}. Quantum
theory---both on complex and real Hilbert spaces---satisfies
purification {\color{black} as do} Spekkens' toy model
\cite{Disilvestro,spekkens2007evidence} 
purification distinct from quantum theory include fermionic quantum
theory \cite{Fermionic1,Fermionic2}, a superselected version of
quantum theory known as double quantum theory
\cite{chiribella2016purity}, and a recent extension of classical
theory to the theory of coherent $d$-level systems, or codits
\cite{chiribella2015entanglement}.  

Pure states $\{|s_i)\}_{i=1}^n$ are called \emph{perfectly
  distinguishable} if there exists a measurement, corresponding to
effects $\{(e_j|\}_{j=1}^n$, with the property that
$(e_j|s_i)=\delta_{ij}$ for all $i,j$.
\begin{define}[Strong symmetry \cite{barnum2014higher,muller2012structure}]
A theory satisfies \emph{strong symmetry} if, for any two $n$-tuples
of pure and perfectly distinguishable states
$\{|\rho_i)\},\{|\sigma_i)\},$ there exists a reversible
transformation $T$ such that $T|\rho_i)=|\sigma_i)$ for $i=1,\dots,n$.
\end{define}

Although complex and real quantum theory satisfy all of the above
principles, double quantum theory and codit theory do not satisfy
strong symmetry.  Whether any other theories satisfy all of the
principles is an important open question.

The following consequences of the above principles, proved in
\cite{lee2016generalised}, will be required to define oracles in
section~\ref{section:oracles}.
\begin{define}
Given a set of pure and perfectly distinguishable states $\{|i)\}$,
and a set of transformations $\{T_i\}$, define a controlled
transformation $C\{T_i\}$ as one that satisfies:
\begin{equation}\label{Control}
\begin{tikzpicture}[scale=1.5]
	\begin{pgfonlayer}{nodelayer}
		\node [style=cpoint] (0) at (0, 1) {$i$};
		\node [style=none] (1) at (1, 1.5) {};
		\node [style=none] (2) at (2.5, 1.5) {};
		\node [style=none] (3) at (1, -0.75) {};
		\node [style=none] (4) at (2.5, -0.75) {};
		\node [style=none] (5) at (1, 1) {};
		\node [style=none] (6) at (2.5, 1) {};
		\node [style=none] (7) at (1, -0.25) {};
		\node [style=none] (8) at (2.5, -0.25) {};
		\node [style=cpoint] (9) at (0, -0.25) {$\sigma$};
		\node [style=none] (10) at (3.5, -0.25) {};
		\node [style=none] (11) at (3.5, 1) {};
		\node [style=none] (12) at (1.75, 1) {$C$};
		\node [style=none] (13) at (5, 0.5) {$=$};
		\node [style=cpoint] (14) at (6.5, 1) {$i$};
		\node [style=none] (15) at (9.5, 1) {};
		\node [style=cpoint] (16) at (6.5, -0.25) {$\sigma$};
		\node [style={small box}] (17) at (8, -0.25) {$T_i$};
		\node [style=none] (18) at (9.5, -0.25) {};
		\node [style=none] (19) at (1.75, -0.25) {$\{T_i\}$};
		\node [style=none] (20) at (11.5, -0) {$\forall i, |\sigma)$};
		\node [style=none] (21) at (12.5, -0) {};
	\end{pgfonlayer}
	\begin{pgfonlayer}{edgelayer}
		\draw (0) to (5.center);
		\draw (9) to (7.center);
		\draw (6.center) to (11.center);
		\draw (8.center) to (10.center);
		\draw (1.center) to (3.center);
		\draw (3.center) to (4.center);
		\draw (4.center) to (2.center);
		\draw (2.center) to (1.center);
		\draw (16) to (17);
		\draw (17) to (18.center);
		\draw (14) to (15.center);
	\end{pgfonlayer}
\end{tikzpicture} \end{equation}
The top system and lower systems are referred to as the \emph{control} and \emph{target} respectively.
\end{define}
Note that classically-controlled transformations---those in which the
control is measured and, conditioned on the outcome, a transformation
is applied to the target---exist in any causal theory with sufficiently
many 
distinguishable states \cite{chiribella2010probabilistic}. However,
such transformations are in general not reversible
\cite{lee2016generalised}.
\begin{thm}[\cite{lee2016generalised} Theorem 2] \label{Reversible-Control}
In any theory satisfying i) causality, ii) purification, iii) strong
symmetry, iv) product of pure states is pure, and in which there
exists a set of $n$ pure and perfectly distinguishable states $|i)$
indexed by $i \in \{1,...,n\}$, for any collection of $n$ reversible
transformations $\{T_i\}_{i=1}^n$ there exists a \emph{reversible}
controlled transformation $C\{T_i\}$ in which the $T_i$ are controlled
by the states $|i)$.
\end{thm}
Every controlled unitary transformation in quantum theory has a
\emph{phase kick-back} mechanism \cite{nielsen2010quantum,
  cleve1998quantum}. Such mechanisms form a vital component of most
quantum algorithms \cite{cleve1998quantum}. It was shown in
\cite{lee2016generalised} that a \emph{generalised} phase kick-back
mechanism exists in any theory satisfying the above physical
principles.

\begin{define}\label{def: phase transformation}
A \emph{phase transformation}, relative to a given measurement of effects $(i|$,
 is a transformation $Q$ such that:  
\[\begin{tikzpicture}[scale=1.5]
	\begin{pgfonlayer}{nodelayer}
		\node [style={small box}] (0) at (1, -0) {$Q$};
		\node [style=none] (1) at (5.5, -0) {};
		\node [style=none] (2) at (0, -0) {};
		\node [style=none] (3) at (8, -0) {$\forall i$};
		\node [style=none] (4) at (4, -0) {$=$};
		\node [style=cocpoint] (5) at (2.5, -0) {$i$};
		\node [style=cocpoint] (6) at (6.5, -0) {$i$};
		\node [style=none] (7) at (8.5, -0) {};
	\end{pgfonlayer}
	\begin{pgfonlayer}{edgelayer}
		\draw (5) to (0);
		\draw (0) to (2.center);
		\draw (6) to (1.center);
	\end{pgfonlayer}
\end{tikzpicture} \]
\end{define}

\begin{thm}[\cite{lee2016generalised} Lemma 2] \label{Generalised-Kick-Back}
In a theory satisfying Causality, Strong Symmetry, and Purification, 
for any set $T_i$ of reversible 
transformations and state $|s)$ such that for all $i$ $T_i|s)=|s)$, there
exists a reversible transformation $Q$ such that
\begin{equation}\label{KB}
\begin{tikzpicture}[scale=1.5]
	\begin{pgfonlayer}{nodelayer}
		\node [style=cpoint] (0) at (0, -0.25) {$s$};
		\node [style=none] (1) at (1.5, 1) {$C$};
		\node [style=cpoint] (2) at (6, 1) {$\sigma$};
		\node [style=none] (3) at (2.25, 1.5) {};
		\node [style=none] (4) at (9, -0.25) {};
		\node [style={small box}] (5) at (7.5, 1) {$Q$};
		\node [style=none] (6) at (0.75, -0.75) {};
		\node [style=none] (7) at (10.5, -0) {$\forall |\sigma)$};
		\node [style=none] (8) at (0.75, 1.5) {};
		\node [style=none] (9) at (0.75, -0.25) {};
		\node [style=none] (10) at (1.5, -0.25) {$\{T_i\}$};
		\node [style=none] (11) at (4.5, 0.5) {$=$};
		\node [style=none] (12) at (9, 1) {};
		\node [style=none] (13) at (3, 1) {};
		\node [style=none] (14) at (2.25, -0.75) {};
		\node [style=none] (15) at (0.75, 1) {};
		\node [style=cpoint] (16) at (6, -0.25) {$s$};
		\node [style=none] (17) at (2.25, -0.25) {};
		\node [style=none] (18) at (3, -0.25) {};
		\node [style=none] (19) at (2.25, 1) {};
		\node [style=cpoint] (20) at (0, 1) {$\sigma$};
		\node [style=none] (21) at (11, -0) {};
	\end{pgfonlayer}
	\begin{pgfonlayer}{edgelayer}
		\draw (20) to (15.center);
		\draw (8.center) to (6.center);
		\draw (6.center) to (14.center);
		\draw (14.center) to (3.center);
		\draw (3.center) to (8.center);
		\draw (0) to (9.center);
		\draw (17.center) to (18.center);
		\draw (19.center) to (13.center);
		\draw (2) to (5);
		\draw (16) to (4.center);
		\draw (5) to (12.center);
	\end{pgfonlayer}
\end{tikzpicture} \end{equation}
where $Q$ is a phase transformation.  Moreover, every phase
transformation can arise via such a generalised phase kick-back
mechanism.
\end{thm}

\subsection{Post-quantum interference}

In the sections that follow, we will connect post-quantum, or
higher-order, probabilistic interference to post-quantum computation;
investigating whether ``more'' interference implies ``more''
computational power. The definition of higher-order interference that
we present here takes its motivation from the set-up of multi-slit
interference experiments. In such experiments a particle (a photon or
electron, say) passes through slits in a physical barrier and is
detected at a screen placed behind the barrier. By blocking some (or
none) of the slits and repeating the experiment many times, one can
build up an interference pattern on the screen.  The ``intensity'' of
the pattern in a small area of the screen is proportional to the
probability that the particle arrives there.  Informally, a theory has
``$n$th order interference'' if one can generate interference patterns
in an $n$-slit experiment which cannot be created in any experiment
with only $m$ slits, for all $m<n$.

More precisely, this means that the interference pattern created on the screen cannot be written as a particular linear combination of the interference patterns generated when different subsets of slits are open and closed. In the standard two slit experiment, quantum interference corresponds to the statement that the interference pattern can't be written as the sum of single slit patterns:
 \[\begin{tikzpicture}[scale=.6]
	\begin{pgfonlayer}{nodelayer}
		\node [style=none] (0) at (0, 1.25) {};
		\node [style=none] (1) at (0, 1.25) {};
		\node [style=none] (2) at (0, -1.25) {};
		\node [style=none] (3) at (0, -1.75) {};
		\node [style=none] (4) at (0, -3) {};
		\node [style=none] (5) at (0, 1.75) {};
		\node [style=none] (6) at (0, 3) {};
		\node [style=none] (7) at (2, -0) {$\neq$};
		\node [style=none] (8) at (4, 1.25) {};
		\node [style=none] (9) at (4, 1.25) {};
		\node [style=none] (10) at (4, -1.25) {};
		\node [style=none] (11) at (4, -1.75) {};
		\node [style=none] (12) at (4, -3) {};
		\node [style=none] (13) at (4, 1.75) {};
		\node [style=none] (14) at (4, 3) {};
		\node [style=none] (15) at (4, 2) {};
		\node [style=none] (16) at (4, 1) {};
		\node [style=none] (17) at (6, -0) {$+$};
		\node [style=none] (18) at (8, 1.25) {};
		\node [style=none] (19) at (8, 1.25) {};
		\node [style=none] (20) at (8, -1.25) {};
		\node [style=none] (21) at (8, -1.75) {};
		\node [style=none] (22) at (8, -3) {};
		\node [style=none] (23) at (8, 1.75) {};
		\node [style=none] (24) at (8, 3) {};
		\node [style=none] (25) at (8, -2) {};
		\node [style=none] (26) at (8, -0.9999998) {};
	\end{pgfonlayer}
	\begin{pgfonlayer}{edgelayer}
		\draw [style=block] (15.center) to (16.center);
		\draw [style=block] (25.center) to (26.center);
		\draw [style=slit] (1.center) to (2.center);
		\draw [style=slit] (6.center) to (5.center);
		\draw [style=slit] (3.center) to (4.center);
		\draw [style=slit] (9.center) to (10.center);
		\draw [style=slit] (14.center) to (13.center);
		\draw [style=slit] (11.center) to (12.center);
		\draw [style=slit] (19.center) to (20.center);
		\draw [style=slit] (24.center) to (23.center);
		\draw [style=slit] (21.center) to (22.center);
	\end{pgfonlayer}
\end{tikzpicture} \]
It was first shown by Sorkin \cite{sorkin1994quantum,sorkin1995quantum} that---at least for ideal experiments \cite{sinha2015superposition}---quantum theory is limited to the $n=2$ case. That is, the interference pattern created in a three---or more---slit experiment \emph{can} be written in terms of the two and one slit interference patterns obtained by blocking some of the slits. 
 Schematically:
\[\begin{tikzpicture}[scale=.65]
	\begin{pgfonlayer}{nodelayer}
		\node [style=none] (0) at (0, -0.25) {};
		\node [style=none] (1) at (0, 0.25) {};
		\node [style=none] (2) at (0, 1.25) {};
		\node [style=none] (3) at (0, -1.25) {};
		\node [style=none] (4) at (0, -1.75) {};
		\node [style=none] (5) at (0, 1.75) {};
		\node [style=none] (6) at (0, 3) {};
		\node [style=none] (7) at (0, -3) {};
		\node [style=none] (8) at (4, -1.25) {};
		\node [style=none] (9) at (4, -1.75) {};
		\node [style=none] (10) at (4, -0.25) {};
		\node [style=none] (11) at (4, -3) {};
		\node [style=none] (12) at (4, 1.75) {};
		\node [style=none] (13) at (4, 3) {};
		\node [style=none] (14) at (4, 1.25) {};
		\node [style=none] (15) at (4, 0.25) {};
		\node [style=none] (16) at (12, -3) {};
		\node [style=none] (17) at (8, 1.75) {};
		\node [style=none] (18) at (8, -3) {};
		\node [style=none] (19) at (12, 1.25) {};
		\node [style=none] (20) at (12, -0.25) {};
		\node [style=none] (21) at (8, 3) {};
		\node [style=none] (22) at (8, -0.25) {};
		\node [style=none] (23) at (12, -1.75) {};
		\node [style=none] (24) at (8, 1.25) {};
		\node [style=none] (25) at (12, 3) {};
		\node [style=none] (26) at (8, 0.25) {};
		\node [style=none] (27) at (12, 0.25) {};
		\node [style=none] (28) at (8, -1.75) {};
		\node [style=none] (29) at (8, -1.25) {};
		\node [style=none] (30) at (12, -1.25) {};
		\node [style=none] (31) at (12, 1.75) {};
		\node [style=none] (32) at (20, -1.75) {};
		\node [style=none] (33) at (24, 3) {};
		\node [style=none] (34) at (24, -1.25) {};
		\node [style=none] (35) at (16, -3) {};
		\node [style=none] (36) at (20, -0.2500001) {};
		\node [style=none] (37) at (20, -1.25) {};
		\node [style=none] (38) at (20, 3) {};
		\node [style=none] (39) at (24, 0.2500001) {};
		\node [style=none] (40) at (16, -1.75) {};
		\node [style=none] (41) at (16, -1.25) {};
		\node [style=none] (42) at (24, -0.2500001) {};
		\node [style=none] (43) at (24, 1.75) {};
		\node [style=none] (44) at (16, -0.2500001) {};
		\node [style=none] (45) at (16, 0.2500001) {};
		\node [style=none] (46) at (20, 1.75) {};
		\node [style=none] (47) at (20, 0.2500001) {};
		\node [style=none] (48) at (20, 1.25) {};
		\node [style=none] (49) at (20, -3) {};
		\node [style=none] (50) at (24, -1.75) {};
		\node [style=none] (51) at (24, -3) {};
		\node [style=none] (52) at (24, 1.25) {};
		\node [style=none] (53) at (16, 1.25) {};
		\node [style=none] (54) at (16, 3) {};
		\node [style=none] (55) at (16, 1.75) {};
		\node [style=none] (56) at (2, -0) {$=$};
		\node [style=none] (57) at (6, -0) {$+$};
		\node [style=none] (58) at (10, -0) { $+$};
		\node [style=none] (59) at (14, -0) { $-$};
		\node [style=none] (60) at (18, -0) { $-$};
		\node [style=none] (61) at (22, -0) { $-$};
		\node [style=none] (62) at (4, 1) {};
		\node [style=none] (63) at (4, 2) {};
		\node [style=none] (64) at (8, 0.4999999) {};
		\node [style=none] (65) at (8, -0.4999999) {};
		\node [style=none] (66) at (12, -0.9999998) {};
		\node [style=none] (67) at (12, -2) {};
		\node [style=none] (68) at (16, 2) {};
		\node [style=none] (69) at (16, 1) {};
		\node [style=none] (70) at (20, 2) {};
		\node [style=none] (71) at (20, 1) {};
		\node [style=none] (72) at (20, -0.9999998) {};
		\node [style=none] (73) at (20, -2) {};
		\node [style=none] (74) at (24, -2) {};
		\node [style=none] (75) at (24, -0.7500001) {};
		\node [style=none] (76) at (24, -0.9999998) {};
		\node [style=none] (77) at (16, -0.4999999) {};
		\node [style=none] (78) at (16, 0.4999999) {};
		\node [style=none] (79) at (24, -0.4999999) {};
		\node [style=none] (80) at (24, 0.4999999) {};
	\end{pgfonlayer}
	\begin{pgfonlayer}{edgelayer}
		\draw [style=block] (62.center) to (63.center);
		\draw [style=block] (65.center) to (64.center);
		\draw [style=block] (67.center) to (66.center);
		\draw [style=block] (69.center) to (68.center);
		\draw [style=block] (71.center) to (70.center);
		\draw [style=block] (73.center) to (72.center);
		\draw [style=block] (74.center) to (76.center);
		\draw [style=block] (77.center) to (78.center);
		\draw [style=block] (79.center) to (80.center);
		\draw [style=slit] (5.center) to (6.center);
		\draw [style=slit] (1.center) to (2.center);
		\draw [style=slit] (3.center) to (0.center);
		\draw [style=slit] (7.center) to (4.center);
		\draw [style=slit] (12.center) to (13.center);
		\draw [style=slit] (15.center) to (14.center);
		\draw [style=slit] (8.center) to (10.center);
		\draw [style=slit] (11.center) to (9.center);
		\draw [style=slit] (17.center) to (21.center);
		\draw [style=slit] (26.center) to (24.center);
		\draw [style=slit] (29.center) to (22.center);
		\draw [style=slit] (18.center) to (28.center);
		\draw [style=slit] (31.center) to (25.center);
		\draw [style=slit] (27.center) to (19.center);
		\draw [style=slit] (30.center) to (20.center);
		\draw [style=slit] (16.center) to (23.center);
		\draw [style=slit] (55.center) to (54.center);
		\draw [style=slit] (45.center) to (53.center);
		\draw [style=slit] (41.center) to (44.center);
		\draw [style=slit] (35.center) to (40.center);
		\draw [style=slit] (46.center) to (38.center);
		\draw [style=slit] (47.center) to (48.center);
		\draw [style=slit] (37.center) to (36.center);
		\draw [style=slit] (49.center) to (32.center);
		\draw [style=slit] (43.center) to (33.center);
		\draw [style=slit] (39.center) to (52.center);
		\draw [style=slit] (34.center) to (42.center);
		\draw [style=slit] (51.center) to (50.center);
	\end{pgfonlayer}
\end{tikzpicture} \]
The terms with minus signs in the above correct for over-counting of the open
slits. If a theory does not have $n$th order interference then one can
show it will not have $m$th order interference, for any $m>n$
\cite{sorkin1994quantum}.  Therefore, one can classify theories
according to their maximal order of interference, $k$. For example
quantum theory lies at $k=2$ and classical theory at $k=1$.

Consider the state of the particle just before it passes through the
slits. For every slit, there should exist states such that the
particle would definitely be found at that slit, if one were to measure
it. Mathematically, this means that there exists a face\footnote{A
  face is a convex set with the property that if $px+(1-p)y$, for some
  $p\in(0,1)$, is an element then $x$ and $y$ are also both elements.}
\cite{barnum2014higher} of the state space, such that all states in
this face give unit probability for the ``yes'' outcome of the two
outcome ``is the particle at this slit?'' measurement. These faces
will be labelled $F_i$, one for each of the $n$ slits $i\in\{1,\dots,
n\}$. As the slits should be perfectly distinguishable, the faces
associated to the slit should be mutually orthogonal. This can be
achieved by letting the slits be in one-to-one correspondence with a
set of pure and perfectly distinguishable states.

One can additionally ask coarse grained questions of the form ``Is the
particle found among a certain subset of slits, rather than somewhere
else?''. The set of states that give outcome ``yes'' with probability
one must contain all the faces associated with each slit in the
subset. Hence the face associated to the subset of slits
$I\subseteq\{1,\dots, n\}$ is the smallest face containing each face
in this subset, $F_I:=\bigvee_{i\in I} F_i$. That is, $F_I$ is the face
generated by the pure and perfectly distinguishable states identified
by the subset $I.$ The face $F_I$ contains all those states which can
be found among the $I$ slits. The experiment is ``complete'' if all
states in the state space (of a given type $A$) can be found among
some subset of slits. That is, if $F_{12\cdots n}=\mathrm{St(A)}$.

Higher-order interference was initially formalised by Rafael Sorkin in
the framework of Quantum Measure Theory \cite{sorkin1994quantum} but
has more recently been adapted to the setting of generalised
probabilistic theories in
\cite{ududec2011three,ududec2012perspectives,barnum2014higher,lee2016generalised,lee2016higher}. A
straightforward translation to this setting describes the order of
interference in terms of probability distributions corresponding to
interference patterns generated in the different experimental setups
(which slits are open, etc.)
\cite{lee2016generalised,lee2016higher}. However, given the principles
imposed in the previous section, it is possible to define physical
transformations that correspond to the action of opening and closing
certain subsets of slits. In this case, there is a more convenient
(and equivalent \cite{barnum2014higher}, given our principles)
definition in terms of such transformations (such a definition was
also used in \cite{ududec2011three, ududec2012perspectives}).

Given $N$ slits, labelled $1, \dots, N$, these transformations will be
denoted $P_I$, where $I \subseteq \{1, \dots, N\}$ corresponds to the
subset of slits which are not closed. In general one expects that $P_I
P_J = P_{I\cap J}$, as only those slits belonging to both $I$ and $J$
will not be closed by either $P_I$ or $P_J$. This intuition suggests
that these transformations should correspond to projectors
(i.e. idempotent transformations $P_IP_I=P_I$). Given the principles
imposed in this paper, this is indeed the case.
\begin{thm} \label{existence of projector}
In any theory satisfying the principles introduced in the previous
section, the projector onto a face generated by a subset of pure and
perfectly distinguishable states is an allowed transformation in the
theory.  If $F$ and $G$ are faces generated by
subsets of the same pure and perfectly distinguishable set
of states, one has $P_FP_G=P_{F\cap G}$.
\end{thm}

The proof of theorem~\ref{existence of projector} is presented in appendix~\ref{Proof of projector}. Given this structure, one can define the maximal order of interference as follows \cite{barnum2014higher,lee2016deriving}.
\begin{define}
A theory satisfying the principles imposed in this section has maximal order of interference $k$ if, for any $N \geq k$, one has:
\[\mathds{1}_N = \sum_{{ \begin{array}{c} I\subseteq \mathbf{N} \\ |I|\leq k \end{array}}}\mathcal{C}\left(k,|I|,N\right)P_I\]
where $\mathds{1}_N $ is the identity on a system with $N$ pure and
perfectly distinguishable states and
\[\mathcal{C}\left(k,|I|,N\right):=(-1)^{k-|I|}\left(\begin{array}{c}N-|I|-1\\ k-|I|\end{array}\right)\]
\end{define}

The factor $\mathcal{C}\left(k,|I|,N\right)$ in the above definition
corrects for the overlaps that occur when different combinations of
slits are open and closed. For $k=N$, the above reduces to the
expected expression $\mathds{1}_h=P_{\{1,...,k \}}$, that is, the
identity is given by the projector with all slits open. The case
$N=k+1$ corresponds to $\mathcal{C}\left(k,|I|,k+1\right) =
(-1)^{k-|I|}$, corresponding to the situation depicted in the above
diagrams, as well as the one most commonly discussed in the literature
\cite{sorkin1994quantum,ududec2011three}.

Instead of working directly with these physical projectors, it is mathematically convenient to work with the (generally) unphysical transformations corresponding to projecting onto the ``coherences'' of a state. Consider the example of a qutrit in quantum theory, the projector $P_{\{0,1\}}$ projects onto a two dimensional subspace:
\[ P_{\{0,1\}}::\left(\begin{array}{ccc} \rho_{00} & \rho_{01} & \rho_{02} \\ \rho_{10} & \rho_{11} & \rho_{12} \\\rho_{20} & \rho_{21} & \rho_{22}   \end{array}\right)\mapsto \left(\begin{array}{ccc} \rho_{00} & \rho_{01} & 0 \\ \rho_{10} & \rho_{11} & 0 \\0 & 0 & 0   \end{array}\right)\]
whilst the coherence-projector $\omega_{\{0,1\}}$ projects only onto the coherences in that two dimensional subspace:
\[ \omega_{\{0,1\}}::\left(\begin{array}{ccc} \rho_{00} & \rho_{01} & \rho_{02} \\ \rho_{10} & \rho_{11} & \rho_{12} \\\rho_{20} & \rho_{21} & \rho_{22}   \end{array}\right)\mapsto \left(\begin{array}{ccc} 0 & \rho_{01} & 0 \\ \rho_{10} & 0 & 0 \\0 & 0 & 0   \end{array}\right).\]
That is, $\omega_{\{0,1\}}$ corresponds to the linear combination of
projectors: $P_{\{0,1\}}-P_{\{0\}}-P_{\{1\}}$.

There is a coherence-projector $\omega_I$ for each subset of slits $I \subseteq \{1,\dots, N\}$, defined in terms of the physical projectors:
$$\omega_I:=\sum_{\tilde{I}\subseteq I}(-1)^{|I|+|\tilde{I}|}P_{\tilde{I}}.$$ These have the following useful properties, which were proved in \cite{lee2016deriving,barnum2014higher}.
\begin{lem} \label{lemma: decomposition of the identity into coherences}
An equivalent definition of the maximal order of interference, $k$, is: $$\mathds{1}_N=\sum_{I,|I|=1}^{k}\omega_I, \text{ for all } N \geq k.$$
\end{lem}
The above lemma implies that any state (indeed, any vector in the vector space generated by the states) in a theory with maximal order of interference $k$ can be decomposed in a form reminiscent of a rank $k$ tensor:
\begin{equation} \label{coherence decomposition}
|s)=\sum_{I,|I|=1}^k \omega_I|s)=\sum_{I,|I|=1}^k |s_I).
\end{equation}
This decomposition can be thought of as a generalised superposition,
as it manifestly describes the coherences between different subsets of
perfectly distinguishable states (the analogue of a basis in quantum
theory) present in a given state.  This will be important in
discussing the power of oracle queries in the following section, since
it allows oracles to be queried not just on states corresponding to 
definite inputs or probabilistic mixtures
of them, but on superpositions of them.

 
\section{Oracles}  \label{section:oracles}

In classical computation, an \emph{oracle} is usually defined as a
total function $f: S \rightarrow T$ from a finite or (more usually)
countably infinite set $S$ to a finite set $T$.  Most commonly, we
have $f:\mathbb{N}\rightarrow\{0,1\}$, or $f: \{0,1\}^* \rightarrow
\{0,1 \}$.  These last are essentially equivalent since the set
$\{0,1\}^*$ of finite binary strings can be identified with
$\mathbb{N}$ via the usual binary encoding.  The string $x$ is said to
be \emph{in} an oracle $O$ if $f(x)=1$, hence oracles can decide
membership in a language (defined as a subset of the set of finite
binary strings).  In a classical oracle model of computation, some
baseline computational model, e.g. a circuit or Turing machine model,
is augmented by the ability to ``query'' the oracle, i.e. obtain the
value of $f$ on one of its inputs.  A query is assigned some cost in
units commensurate with those of the baseline model, and multiple
queries may be made in the course of a computation, on inputs provided
in terms of the baseline model (normally, bit strings on a tape or in
some set of registers).  Oracle outputs are also provided in terms of
the baseline model, and may be further processed by means of the
baseline model's resources and/or as input to additional queries.
Sometimes a model is considered in which the resources of the baseline
model are taken to be free, and the \emph{only} cost is the number of
queries to the oracle; this is usually termed a \emph{query model}.
   
In quantum computation oracle queries to a function $f: \{0,1\}^* \rightarrow \{0,1\}$ are usually represented by a family
$\{G_n\}$ of quantum gates, one for each length $n$ of the ``input''
string $x$.\footnote{Other models of quantum oracle queries have been
  investigated, but this one is by far the most common.}  Each $G_n$
is a unitary transformation acting on $n+1$ qubits, whose effect on
the computational basis is in general given by
\begin{equation}\label{quantum oracle version 1}
G_n\ket{x,a}= \ket{x,a\oplus{f_n(x)}}
\end{equation} for all $x\in\{0,1\}^n$ and
$a\in\{0,1\}$, where $f_n$ are a family of Boolean functions that
represent the specific oracle under consideration.  Since the family
$f_n$ determines (and is determined by) a function $f: \{0,1\}^*
\rightarrow \{0,1\}$, where $f_n$ is $f$'s restriction to inputs of length $n$, 
a quantum oracle may be thought of as a
``coherent'' version of the corresponding classical oracle $f$; we
call it a ``quantum oracle for $f$''.  Slightly more generally, one
could define a quantum oracle (``for $f$'') as a family (indexed by $|x|$) 
of controlled unitary
transformations which, when queried by state $\ket{x}$ in the control
register, applies a unitary---chosen from a set of two unitaries
according to the value $f_n(x)$---to the target register. A specific
example of a quantum oracle of this sort is the following controlled
unitary:
\begin{equation} \label{Quantum-Oracle}
U_f= \ket{0}\bra{0}\otimes Z^{f(0)} + \ket{1}\bra{1}\otimes Z^{f(1)},
\end{equation}
with $Z$ the Pauli $Z$ matrix, $f:\{0,1\}\rightarrow\{0,1\}$ a
function encoding some decision problem and $Z^{0}:=\mathds{I}$.  (If
we had used the Pauli $X$ matrix, in place of $Z$, this oracle would
be of the type described by Eq. \ref{quantum oracle version 1}.

As was briefly mentioned in \cite{lee2016generalised}, the results of
theorem~\ref{Reversible-Control} provide a way to define computational
oracles in any theory satisfying our assumptions. 

\begin{define}\label{def:oracle}
In an theory satisfying our assumptions, an oracle for a decision
problem $f: S \rightarrow T$, with $S$ and $T$ finite sets, is a
reversible controlled transformation\footnote{There could be many
  distinct transformations that have the same behaviour on a set of
  control states. As long as one fixes which transformation
  corresponds to the oracle, this is not a problem.} $C\{T_i\}$ where
the set of transformations $\{T_i \}$ being controlled depend on $i
\in S$ \emph{only} through the value $f(i)$.  If $S = \{A\}^*$, the
countably infinite set of strings from a finite alphabet $A$, 
then an oracle for $f$ is defined to be a family
$G_n$, indexed by the length $n$ of strings, of such controlled
transformations, one for each $f_n$, where $f_n$ is defined to be $f$
restricted to strings of length $n$.  If $T = \{0,1\}$, this is
a decision problem.
\end{define}

In quantum theory there is an equivalent view of oracles in terms of
phase transformations.  This can be seen as a result of the phase
kick-back algorithm \cite{cleve1998quantum,nielsen2010quantum}. In the
quantum example above, the phase kick-back for $U_f$ amounts to
{\color{black} first rewriting $U_f$ as:}
\begin{equation} \label{Quantum-kick-back}
U_f=\mathds{I}\otimes\ket{0}\bra{0}+Z^{f(0)\oplus f(1)}\otimes\ket{1}\bra{1}.
\end{equation}
{\color{black} Inputting $\ket{1}$ into the second qubit results in a
  `kicked-back' phase of $Z^{f(0)\oplus f(1)}$ on the first qubit. The
  value of $f(0)\oplus f(1)$ can then be measured by preparing the
  first qubit in the state $\ket{+}$ and then} measuring
{\color{black}it} in the $\{\ket{+},\ket{-}\}$ basis. {\color{black}
  This provides} the value $f(0)\oplus f(1)$ in a single query of the
oracle---a feat impossible on a classical computer
\cite{meyer2011uselessness}.

In our more general setting, an analogue of the above holds via
Theorem \ref{Generalised-Kick-Back}. That is, in theories satisfying
our assumptions, as the transformations $T_{i}$ depend on the value
$f(i)$, so {\color{black} too can} the controlled transformation and
the kicked-back phase.  That is, in {\color{black} theories
with non-trivial phases, i.e. non-classical theories, the phase
kick-back of an oracle can encode} information about the value $f(i)$
for all $i$.  Indeed, it is also the case that if one has available as
one circuit element the generalised phase kick-back transformation
constructed out of the controlled transformation $C{T_i}$ one can
construct a circuit for $C{T_i}$ \cite{lee2016generalised}.
Hence---as in the quantum example above---from the point of view of
querying the oracle, one can reduce all considerations involving the
controlled transformation to those involving the phase kick-back
transformation, which shall be denoted $\mathcal{O}_f$.  (This justifies
our use of phase transformations as oracles in Definition 
\ref{def: oracle system} below.)


As was shown in section~\ref{framework}, all states in theories
satisfying our principles can be decomposed as $|s)=\sum_I |s_I)$,
with $I\subseteq\{1,\dots, n\}$, where $\{1,\dots, n\}$ labels the set
of pure and perfectly distinguishable states defining the action of a
give oracle.  Hence, oracles can not only be queried using a set of
pure and perfectly distinguishable states, but also using generalised
superposition states---those with non-trivial coherences between
different subsets of slits. {\color{black} In fact, the quantum
  speed-up in the above example came precisely from the fact that one
  can query in superposition, hence extracting the value $f(0)\oplus
  f(1)$ in a single query.}  To ensure that answers to hard to solve
problems are not smuggled into the definition of oracles in
generalised theories, we must put conditions on which phase
transformations correspond to `reasonable' oracles.

We remark that in some definitions of oracle, the possibility of a 
\emph{null query} is included.  This is an input to the oracle transformation
conditional on which nothing happens to the target register.  Although we 
did not explicitly include the possibility above, we could define an oracle
for $f: S \rightarrow T$ with the possibility of a null query, as above but with
an additional distinguishable state of the control register, indexed by some symbol
(say $\bullet$) not in $S$, and with $T_{\bullet} = \mathrm{id}$, the identity 
transformation.  Our results still hold for
this notion of oracle (which can be viewed as a special case of an oracle of the 
type defined above, for the slight extension of the function $f$ to have one more 
input, on which it takes a fixed value). 

{\color{black}
\begin{define}[Oracle system] \label{def: oracle system}
A \emph{system of oracles} for a family $\mathcal{C}$ of functions $S
\rightarrow T$ is defined as a family of phase transformations (which
we call ``oracles'' because they are a particular case of the oracles
of Definition \ref{def:oracle}) $\{\mathcal{O}_f\}_{f \in
  \mathcal{C}}$ such that whenever $f(i)=g(i)$ for all $i\in I$, the
oracles corresponding to $f$ and $g$
satisfy $$\mathcal{O}_f|s_I)=\mathcal{O}_g|s_I)$$ for all $|s_I)$ of
the form $\omega_I |s)$ (for arbitrary $|s)$.
\end{define}

An equivalent, and perhaps more intuitively motivated, definition
substitutes the condition `` for all states $|s_I)$ such that $|s_I) =
P_I |s_I)$, for the condition ``for all $|s_I)$ of the form $|s_I) =
\omega_I|s)$...''  This ensures one cannot learn about the value
$f(j)$ when querying using a state with no probability of being found
in $|j)$.}  That is, $\mathcal{O}_f$ and $\mathcal{O}_g$ act
identically on states in the face determined by a subset of inputs on
which $f$ and $g$ have the same value, so that we cannot, for example,
just write \emph{arbitrary} information about which function is being
queried into phase degrees of freedom.

One can schematically represent the problems that can be solved by a
specific computational model with access to an oracle 
using the language of complexity classes. Let $\mathbf{C}$ and
$\mathbf{B}$ be complexity classes, then $\mathbf{C}^{\mathbf{B}}$
denotes the class $\mathbf{C}$ with an oracle for $\mathbf{B}$ (see
\cite{papadimitriou2003computational} for formal definitions). We can
think of $\mathbf{C}^{\mathbf{B}}$ as the class of languages decided
by a computation which is subject to the restrictions and acceptance
criteria of $\mathbf{C}$, but allowing an extra kind of computational
step: an oracle for any desired language $\mathcal{L}\in\mathbf{B}$
that may be queried during the computation, where each query counts as
a single computational step.  Here $\mathcal{L}$ is fixed in any
  given computation, though different computations may use
  different $\mathcal{L}$.

A natural question is whether or not having access to an
oracle 
for a particular decision problem which can be
efficiently solved in a given theory provides any more computational
power than just using the efficient algorithm.   If we schematically
denote the class of problems efficiently solvable by a particular
theory \textbf{G} by\footnote{See references
  \cite{lee2015computation,lee2015proofs,landscape} for a rigorous
  definition of this class.} \textbf{BGP}, this question can be
phrased as: ``is \textbf{BGP} closed under \emph{subroutines}''? Here
\textbf{BGP} is the analogue of the well-known class of problems
efficiently solvable by a quantum computer, \textbf{BQP}. Another way
to pose this question is to ask whether
$\textbf{BGP}^\textbf{BGP}=\textbf{BGP}$ for \textbf{G} satisfying our
principles.

There exist complexity classes for which this is probably not the
case, for example, \textbf{NP}\footnote{If one assumes that the
  polynomial hierarchy doesn't collapse.}. But, intuitively, one would
expect it to hold in a sensible physical theory where computation is
performed with circuits.  In such a situation, one might consider it a
kind of conceptual consistency check on one's definition of oracle: it
is not reasonable for an oracle for a problem to give more power than
would be afforded by a circuit for that problem in the model.  A
potential issue arises when one compares the performance of the oracle
implementation to that of the efficient algorithm when both are used
as subroutines in another computational algorithm\footnote{Here, an
  algorithm consists of a poly-size uniform circuit. See
  \cite{lee2015computation} for the formal definitions.}.  As we noted
in Sections \ref{framework} and \ref{section:oracles}, an oracle can
be queried on a superposition of inputs, but one does not normally
query an algorithm for a particular decision problem in superposition
for the purpose of solving that decision problem; one merely prepares the
state corresponding to a particular bit string and uses the algorithm
to determine whether or not that bit string is in the language in
question.\footnote{However, a quantum algorithm for one problem can
  sometimes be used as a subroutine in a quantum algorithm for another
  problem; in this case, the subroutine sometimes \emph{is} run on a
  superposition of inputs.}  For simplicity, we say the efficient
algorithm accepts an input if a measurement of the first outcome
system yields outcome $(0|$ with probability\footnote{This can be
  amplified to $1-2^{-q}$, where $q$ a polynomial in the size of the
  circuit, by running the circuit in parallel a polynomial number of
  times. Again, see \cite{lee2015computation}.} greater than
$2/3$. This is the same acceptance condition imposed in quantum
computation.

{\color{black} We therefore need to know whether} every \textbf{BGP}
algorithm for a decision problem admits a subroutine having the
characteristics of an oracle for that decision problem. Such a result
was proved in the quantum case by Bennett et al. in
\cite{bennett1997strengths}. The following theorem shows that it is
also true for theories satisfying our principles.  (See
\cite{lee2015computation} for the definition of circuit and circuit
family in the GPT context.)

\begin{thm}\label{subroutine}
Consider a theory \textbf{G} which satisfies the principles outlined
in section~\ref{framework}. Given an algorithm (poly-size family of
circuits in $\textbf{G}$), $\{A_{|x|}\}$, for a decision problem in
\textbf{BGP}, one can always construct a family $\{G_{|x|}\}$ of
polynomial-size circuits implementing reversible transformations from
\textbf{G}, which, with high probability (greater than $1 -
2^{-q(|x|)}$, for some polynomial $q$), functions as an oracle for
that particular decision problem (in the sense of 
Definition \ref{def:oracle}). (Here, poly-size means polynomial in
the length $|x|$ of the input $x$, and the family $G_{|x|}$ comprises
a fixed circuit for each input size.)  Schematically, we have
$\textbf{BGP}^\textbf{BGP}=\textbf{BGP}$.
\end{thm}
\proof See Appendix \ref{Proof of subroutine} \endproof

Given our definition of an oracle we can consider how their
computational power depends on the order of interference of the
theory.  

\section{Lower bounds from useless queries} \label{lower bound}

In this section we generalise results of
Ref. \cite{meyer2011uselessness}, in which Meyer and Pommersheim derived a
relation between quantum and classical query complexity lower bounds, 
by introducing the concept of a ``useless'' quantum query to the
setting of GPTs satisfying our principles. They considered
\emph{learning problems} in which one is given an element from a class
of functions with the same domain and range, chosen with some
arbitrary---but known---prior distribution, where the task is to
determine to which specific subclass the chosen function belongs.

More formally we can define a learning problem as follows:
\begin{define}[Learning problems]
Given a set of functions $\mathcal{C}\subseteq \{0,1\}^X$ where $X$ is
some finite set\footnote{One could alternatively consider replacing
  $\{0,1\}$ with a different finite set $Y$. Thus the result proved in
  this section, and the result proved in \cite{meyer2011uselessness},
  also holds in the more general case.}, a partitioning of
$\mathcal{C}$ into disjoint subsets $\mathcal{C}=\bigsqcup_{j\in
  J}\mathcal{C}_j$ labeled by $j\in J$, and a probability measure
$\mu$ over $\mathcal{C}$.  The aim of the learning problem is to
determine, which partition $\mathcal{C}_j$ a particular function $f\in
\mathcal{C}$ belongs to; this is to be done with low \emph{ex ante}
probability of error, with respect to the probability measure $\mu$
with which the function is chosen from among the $\mathcal{C}_j$'s.  A
particular learning problem is therefore defined by the triple,
$(\mathcal{C},\{\mathcal{C}_j\},\mu)$.
\end{define}

One can only access information about the function by querying an
oracle, which, when presented with an element from the domain, outputs
the corresponding element of the range assigned by the chosen
function. Typically one specifies some upper bound to the error
probability, and is interested in the minimal number of queries needed
to ensure that the \emph{ex ante} probability of error is below the
bound, with respect to the measure $\mu$ that gives the ``prior
probability'' of functions $f \in \mathcal{C}$.

Meyer and Pommersheim showed that if $n$ queries to a
classical oracle reveal no information about which function was
chosen\footnote{That is, if the probability that the chosen function
  belongs to a given subclass after $n$ classical queries is the same
  as the known prior probability with which it was originally chosen.}
then neither do $n/2$ queries to a quantum oracle. Hence
$\lceil{n/2}\rceil +1$ quantum queries constitute a lower bound.

Many important query problems are examples of learning problems. For
instance, \texttt{PARITY}, a generalisation of the special case of
Deutsch's problem where the input to $f$ is a bit,
\cite{nielsen2010quantum} which asks for the parity of a
function\footnote{i.e. the value $f(1)\oplus\cdots\oplus f(N) \text{
    mod }2$} $f:\{1,\dots,N\}\rightarrow\{0,1\}$ can be written as a
learning problem.  Indeed, partition the class of all such functions
into two subclasses, one in which all functions have parity $0$ and
the other $1$, and choose the function with a prior probability of
$1/2$. In this case, $N-1$ classical queries do not provide any
information about the parity, hence at least $\lceil{(N-1)/2}\rceil
+1$ quantum queries are needed to solve the problem.  In fact 
$\lceil{(N-1)/2}\rceil +1$ quantum queries are also
sufficient\footnote{$\lceil{(N-1)/2}\rceil +1$ applications of the
  solution to Deutsch's problem.}.

In this section we generalise Meyer and Pommersheim's result to the
case of oracle queries in the generalised probabilistic theory
framework presented in the previous section. We prove that if $n$
queries to a classical oracle reveal no information about which
function was chosen then neither do $n/k$ queries in a generalised
theory satisfying the principles introduced in section~\ref{framework}
and which has maximal order of interference $k$. Hence a lower bound
to determining the function is $\lceil{n/k}\rceil+1$ queries in
theories with $k$th order interference.  So in the specific
generalisation of Deutsch's problem where we are asked to determine
the parity of a function $f:\{1,\dots,n\}\rightarrow\{0,1\}$,
$\lceil{n/2}\rceil$ quantum queries are needed, but in a theory with
$n$th order interference, the `` no-information'' or
``useless-queries'' bound does not rule out the possibility that the
parity can be determined with a single query.


We can now formally define what it means for $n$ classical queries to be useless \cite{meyer2011uselessness}.

\begin{define}[Useless classical queries \cite{meyer2011uselessness}]
Let $\left(\mathcal{C}, \{\mathcal{C}_j : j\in J\}, \mu \right)$ be a learning problem. $n$ classical queries are said to be \emph{useless}, or to convey no information,  if for any $x_1, \dots x_n\in X$ and $y_1,\dots y_n \in \{0,1\}$ the following holds
$$\mu\left(f\in\mathcal{C}_j \text{ }  |\text{ } f(x_i)=y_i, i=1,\dots,n\right)=\mu\left(f\in\mathcal{C}_j\right), \text{  for all } j\in J.$$
\end{define}

Here expressions like $\mu\left(f\in\mathcal{C}_j \text{ } |\text{ }
f(x_i)=y_i, i=1,\dots,n\right)$ are to be understood simply as
conditional probabilities of events like $f \in \mathcal{C}_j$, 
conditional on events like $f(x_1)= y_1~ \& ~f(x_2) = y_2 ~\& ~ \cdots ~ \& ~ 
f(x_n) = y_n$.


A general $n$-query algorithm in a generalised theory satisfying our
principles corresponds to the following: an arbitrary initial state
$|\sigma)$ is prepared and input to the oracle $\mathcal{O}_f$, the
output state is acted upon by an arbitrary transformation $G_1$
independent of $f$, and the process is repeated. After the $n$th
oracle query, the state is
$$|\rho_f)=G_n\mathcal{O}_f G_{n-1}\cdots G_1\mathcal{O}_f|\sigma).$$
The final step consists of measuring this state with an arbitrary
measurement denoted as $\{(s|\}_{s\in S}$.  The final\footnote{As was
  noted in \cite{meyer2011uselessness}, the final transformation $G_n$
  is unnecessary, as it could be incorporated into the measurement.}
output of the algorithm is given by a map, which is independent of $f$
from the set $S$ indexing the measurement outcome to the set
$J$ indexing the subclasses to which the function could belong.

The probability of outcome $(s|$ being observed in the measurement,
when the unknown function is $f$, is therefore defined to be
$(s|\rho_f)$.  So there is a joint probability distribution,
which we will denote also by the letter $\mu$, over the outcome $s$
and the function $f$:
$$\mu(s,f) = (s|\rho_f)\mu(f).$$
Bayes' rule gives the 
posterior probability that the function was $f$, given the observed
measurement outcome $s$:
$$\label{applying Bayes rule} \mu(f|s) = (s|\rho_f)\mu(f) / \sum_{g
  \in \mathcal{C}} (s|\rho_g)\mu(g).$$ The posterior probability that
$f \in \mathcal{C}_j$ given this outcome is
$$\mu(f \in \mathcal{C}_j|s) = \sum_{f in \mathcal{C}_j} \mu(f|s).$$
Similarly we define $\mu(f \in C_j) = \sum_{f \in \mathcal{C}_j}
\mu(f)$, the prior probability that $f \in C_j$.

 We can now generalise the definition of a useless quantum query from
 \cite{meyer2011uselessness} to the case of generalised theories
 satisfying our principles.

\begin{define}[Useless generalised queries]
Let $\left(\mathcal{C}, \{\mathcal{C}_j \}_{j\in J}, \mu \right)$ be a learning problem. $n$ generalised queries are said to be \emph{useless}, or to convey no information,  if for any $n$ query generalised algorithm with initial state $|\sigma)$, transformations $G_n,\dots, G_1$, and measurement $\{(s|\}_{s\in S}$ the following holds
$$\mu\left(f\in\mathcal{C}_j \text{ }  |\text{ } s\right)= \mu\left(f\in\mathcal{C}_j\right), \text{  for all possible } s\in S, j\in J.$$
\end{define}

We now present our main result, which
generalises Theorem $1$ from \cite{meyer2011uselessness}.
\begin{thm}
Let $\left(\mathcal{C}, \{\mathcal{C}_j : j\in J\}, \mu \right)$ be a
learning problem.  Suppose $kn$ classical queries are useless. Then in
any theory which satisfies our principles and has maximal order of
interference $k$, $n$ generalised queries are useless.
\end{thm}
We present the formal proof below, but first provide a rough sketch
proof. In a theory with $k$th order interference, each state can be
decomposed as in Eq.~(\ref{coherence decomposition}). Hence each state
is explicitly indexed by subsets---of size at most $|I|=k$---of the
set of pure and perfectly distinguished states defining the
oracle. Thus, after a single generalised query, the state is indexed
by the valuation of the chosen function on at most $k$ inputs. After
$n$ generalised queries, it is indexed by $kn$ valuations. Therefore,
a measurement can reveal at most $kn$ valuations of the chosen
function. But, as $kn$ classical queries are useless, it must be that
$n$ generalised queries are also useless. The intuition behind this
result is that, as a given state can have coherence between at most
$k$ basis states, one can use generalised superposition states to
extract at most $k$ valuations of a given function in a single query.

\begin{proof}
Our proof is essentially a slight generalisation of the original quantum one presented in \cite{meyer2011uselessness}. We need to show that the probability of $f$ being in $\mathcal{C}_j$ does not change if outcome $s$ is observed after $n$ queries. That is, we must show
$$\sum_{f\in\mathcal{C}_j} \mu (f\text{ }|\text{ } s) = \mu\left(\mathcal{C}_j\right), \text{ for any } s\in S \text{ and } j\in J.$$
Recalling from the application of Bayes' rule to obtain 
Eq. \ref{applying Bayes rule} we have:
$$
\mu(f|s)= 
\frac{(s|\rho_f)\mu(f)}{\sum_{g\in \mathcal{C}} (s|\rho_g)\mu(g)}
$$
and summing over $f$ in $\mathcal{C}_j$, we have
\begin{equation} \label{proof}
\sum_{f\in\mathcal{C}_j}\mu (f\text{ }|\text{ } s)  = \frac{(s| \sum_{f\in\mathcal{C}_j} \mu(f)|\rho_f)}{(s| \sum_{g\in\mathcal{C}} \mu(g)|\rho_g)}.
\end{equation}
Let's focus on $|\rho_f)$. Given the decomposition in Eq.~(\ref{coherence decomposition}), every state can be written as
$$|\sigma)=\sum_{I,|I|=1}^k\omega_I|\sigma)=: \sum_{I,|I|=1}^k\sigma_I.$$
Now, each $\mathcal{O}_f\left(\sigma_I\right)$ can depend on all $f(i)$ with $i\in I$. By padding out those $I$ with $|I|<k$ with dummy indices, after a single query one can write
$$\mathcal{O}_f |\sigma)=\sum_I \mathcal{O}_f(\sigma_I)=\sum_{T_1} Q_{T_1}\left( f(x_1^1), f(x^2_1),\dots, f(x^k_1)\right),$$
where the second equality is just a relabeling of the terms where $T_1=\{x_1^1, x^2_1,\dots, x^k_1\}$ is the padded version of $I$, and hence  each $Q_{T_1}$ is a vector in the real vector space of states that depends on $f(x_1^1), f(x^2_1),\dots, f(x^k_1)$.
Therefore, after $n$ queries one can write the state as
$$|\rho_f)= \sum_{T_n} Q_{T_n}\left( f(x_1^1),\dots, f(x^1_n), f(x^2_1),\dots, f(x^2_n),\dots, f(x^k_1),\dots,f(x^k_n) \right) $$
Using a change of variables provides
$$\begin{aligned}
&\sum_{f\in\mathcal{C}_j} \mu(f)|\rho_f)= \\
&\sum_{T_n}\sum_{\{y_i^1\},\dots,\{y^k_i\}} \mu\left(f\in\mathcal{C}_j \text{ and } f(x_i^m)=y_i^m, \text{ for } i=1,\dots,n \text{ and } m=1,\dots, k\right) Q_{T_n}\left(y_1^1,\dots y_n^k \right).
\end{aligned}$$
As $kn$ classical queries are useless
$$\begin{aligned}
\mu & \left(f\in\mathcal{C}_j \text{ and } f(x_i^m)=y_i^m, \text{ for } i=1,\dots,n \text{ and } m=1,\dots, k\right)= \\
& \qquad\qquad\qquad\qquad\qquad\qquad\qquad \mu(\mathcal{C}_j) \mu\left( f(x_i^j)=y_i^j, \text{ for } i=1,\dots,n \text{ and } j=1,\dots, k\right). \end{aligned}$$
Inputting this into the above we obtain,
\begin{equation}\label{eq:proofEdit}
\begin{aligned}
&\sum_{f\in\mathcal{C}_j} \mu(f)|\rho_f)= \\
& \mu(\mathcal{C}_j) \sum_{T_n}\sum_{\{y_i^1\},\dots,\{y^k_i\}}\mu\left( f(x_i^j)=y_i^j, \text{ for } i=1,\dots,n \text{ and } j=1,\dots, k\right) Q_{T_n}\left(y_1^1,\dots y_n^k \right).
\end{aligned}\end{equation}
Then. summing over $j\in J$, results in
$$\begin{aligned}
&\sum_{f\in\mathcal{C}} \mu(f)|\rho_f)= \\
&\sum_{T_n}\sum_{\{y_i^1\},\dots,\{y^k_i\}}\mu\left( f(x_i^j)=y_i^j, \text{ for } i=1,\dots,n \text{ and } j=1,\dots, k\right) Q_{T_n}\left(y_1^1,\dots y_n^k \right).
\end{aligned}$$
Substituting this back into Eq.~(\ref{eq:proofEdit}) immediately gives
$$\sum_{f\in\mathcal{C}_j} \mu(f)|\rho_f) = \mu(\mathcal{C}_j)\sum_{f\in\mathcal{C}}\mu(f)|\rho_f).$$
finally, substituting this into Eq.~(\ref{proof}) completes the proof.

\end{proof}

\section{Conclusion}

In this work we have introduced a well-defined oracle model for
generalised probabilistic theories, and shown it to be well-behaved in
the sense given by our subroutine theorem: that an oracle of our type
for a given problem is not more powerful than an algorithm for that
problem, since an algorithm permits high-probability simulation of an
oracle.  This allowed us to compare the computational power imposed by
different physical principles through the lens of query complexity.
Our main result in this regard was to show that the
``zero-information'' lower bound on the number of queries to a quantum
oracle needed to solve certain problems is not optimal in the space of
generalised theories satisfying the principles introduced in
section~\ref{framework}.  Our result highlights the role of
interference in computational advantages in a theory independent
manner, allowing the possibility that ``more interference could permit
more computational power''.

Previous work by the authors in \cite{lee2016deriving} derived
Grover's lower bound to the search problem from simple physical
principles.  The search problem asks one to find a certain ``marked
item'' from among a collection of items in an unordered database. The
only access to the database is through an oracle; when asked if item
$i$ is the marked one, the oracle outputs ``yes'' or ``no''. The
figure of merit in this problem is how the minimum number of queries
required to find the marked item scales with the size of the
database. It was shown---subject to strong assumptions close to those
used in the present paper---that, asymptotically, higher-order
interference does not provide an advantage over quantum theory in this
case. As opposed to the asymptotic behaviour of the number of queries
needed to solve the search problem, the current work was concerned
with whether a fixed number of queries yielded any information about
the solution of a particular query problem, where the problem could be
any of a large class of ``learning problems''.  In this case we were
able to show that the ``useless-queries'' or ``zero-information''
lower bound on the number of queries is lower, the higher the order of
interference in the theory, leaving open the possibility that
higher-order interference could lead to a computational speedup
(although we did not show that such a speedup is achievable).  Note
that a specific oracle model for the search problem was introduced in
\cite{lee2016deriving}. However, this is just a special case of the
general model introduced in section~\ref{section:oracles} of the
current work. Moreover, the subroutine theorem proved here shows that
our general oracle model is well-defined.

Our derivation of query lower bounds raises the question of whether
the physical principles we have discussed are sufficient for the
existence of algorithms which achieve these lower bounds. In the
specific case of the search problem, a quantum search algorithm based
on Hamiltonian simulation, due to Farhi and Gutmann
\cite{FarhiGutmann} and also presented in chapter $6$ of the well known
textbook by Nielsen and Chuang \cite{nielsen2010quantum}, may be more
directly generalisable to theories satisfying our principles than
Grover's original construction \cite{grover1997quantum}. This approach
may also be applicable to many other query algorithms. In the
algorithm as presented in \cite{nielsen2010quantum} they consider a
Hamiltonian $H$ consisting of projectors onto the marked item
$\ket{x}$ and the initial input state
$\ket{\psi}=\alpha\ket{x}+\beta\ket{y}$, with $\ket{y}$ orthogonal to
$\ket{x}$ and $\alpha^2 +\beta^2=1$, respectively. That is, they
consider the Hamiltonian
$H=\ket{x}\bra{x}+\ket{\psi}\bra{\psi}$. Evolving the initial input
state under this Hamiltonian for time $t$ results in
$$\exp(-itH)\ket{\psi}=\cos(\alpha t)\ket{\psi} - i\sin(\alpha t)\ket{x}.$$
Hence, measuring the system in the $\{\ket{x}, \ket{y}\}$ basis at time $t=\pi/2\alpha$ yields outcome $\ket{x}$ with probability one. If the initial state was a uniform superposition over the orthonormal basis containing $\ket{x}$, then the required evolution time is $t=\pi\sqrt{N}/2$, where $N$ is the size of the system (or equivalently, the number of elements in the database being searched).

One might wonder why there is no mention of an oracle in the above
discussion. The oracle comes into play when constructing a quantum
circuit to simulate the above Hamiltonian evolution. As the above
Hamiltonian depends on the marked item, the quantum circuit simulating
it must query the search oracle a number of times proportional to the
evolution time \cite{nielsen2010quantum}. In this specific case, an
efficient Hamiltonian simulation requires $O(\sqrt{N})$ queries to the
oracle, yielding an optimal quantum algorithm (up to constant factors)
for the search problem. Recently, the authors of
\cite{barnum2014higher} have introduced a physical principle, termed
``energy observability'', which implies the existence of a continuous
reversible time evolution and ensures that the generator of any such
evolution---a generalised ``Hamiltonian''---is associated to an
appropriate observable, which is a conserved quantity under the
evolution---the generalised ``energy'' of the evolving system.  Recall
from section~\ref{framework} that the principles we have discussed
were sufficient to ensure that projectors onto arbitrary states
correspond to allowed transformations.  Hence, our previous
principles, together with energy observability as introduced in
\cite{barnum2014higher}, might be sufficient to run the above quantum
search algorithm, hence providing a theory independent description of
an optimal (up to constant factors) search algorithm. Similar
constructions based on Hamiltonian simulation might also show that
theories satisfying the above physical principles can reach the query
lower bounds derived in this paper.

\section*{Acknowledgements}
The authors thank D. Meyer for bringing to their attention his work on useless quantum queries with J. Pommersheim in \cite{meyer2011uselessness}. The authors thank Matty Hoban for useful discussions and J.J. Barry for encouragement while writing the current paper. This work was supported by EPSRC grants through the Controlled Quantum Dynamics Centre for Doctoral Training, and the UCL Doctoral Prize Fellowship. We also acknowledge financial support from the European Research Council (ERC Grant Agreement no 337603), the Danish Council for Independent Research (Sapere Aude) and VILLUM FONDEN via the QMATH Centre of Excellence (Grant No. 10059).  This work
began while the authors were attending the ``Formulating and Finding Higher-order Interference'' workshop at the Perimeter Institute. Research at Perimeter Institute is supported by the Government of Canada through the Department of Innovation, Science and Economic Development
Canada and by the Province of Ontario through the Ministry of Research, Innovation and Science.

\begin{appendices}
\section{Proof of theorem~\ref{existence of projector}} \label{Proof of projector}

This is an adaptation of the proof of theorem 8 from
\cite{barnum2014higher}. Consider a self-dual cone $\mathbf{C}$ with
self-dualising inner product $\left<\cdot,\cdot\right>$. Now consider 
a set of pure and perfectly distinguishable states $\phi_i$ which are
distinguished by the effects $e_i$ such that
$(e_i|\phi_j)=\delta_{ij}$. We can define a face $F$ of $\mathbf{C}$
as the minimal face generated by the set of states $\{\phi_i\}$, we
can moreover define the dual face $F^*:=\left\{x\in \mathbf{C}\ |
\left<x,s\right> \geq 0 \ \ \forall s\in F\right\}$. Appendix A in
\cite{barnum2014higher} shows that if $F=F^*$ then there exists a
positive projector onto the face $F$.

Consider some $t\in F$.  Self-duality of $\mathbf{C}$ implies that
$\left<t,s\right>\geq 0 \ \forall s\in F$ hence, $t\in F^*$ and so
$F\subseteq F^*$. We therefore just need to prove the converse
inclusion and we are done.

To prove this, consider a normalised extremal $x\in F^*$, there must be some $s\in F$ such that $\left<s,x\right>=0$, where moreover, if $\left<s,y\right>=0$ then $y\propto x$. Next we prove two simple results:

i) $s$ is not internal to $F$---assume, for the sake of contradiction, that $s$ is internal. Then $\left<x,t\right>=0 \ \forall t\in F$ so $x=0$ and hence is not normalised.

ii) There exists $t\in F$ such that $s$ and $t$ are perfectly
distinguishable---assume, again to reach a contradiction, that there
is no such $t$. This means that given any pure and perfectly
distinguishing measurement $\{\epsilon_i\}$ that $(\epsilon_i|s)>0$
for all $i$. Due to strong symmetry, any pure effect appears in such a
measurement, therefore $(e|s)>0$ for all pure effects $(e|$, this
suffices for tomography hence $|s)$ is an internal state, in
contradiction with (i).

%

Theorem 1 from \cite{muller2012structure} implies that if $s$ and $t$ are perfectly distinguishable states then $\left<s,t\right>=0$, therefore we know that $t\propto x$ and so $x\in F$. This is true for all extremal normalised $x\in F^*$ it therefore follows from convexity that this is true for all $x\in F^*$ and so we have $F^*\subseteq F$ which concludes the proof.

Hence, projectors $P_F$ onto $F$ are positive transformations. It was shown in \cite{chiribella2010probabilistic} that in any theory satisfying causality, purification and informationally consistent composition, mathematically well-defined transformations are physical, i.e. they are allowed in the theory. Hence projectors $P_F$ are physically allowed transformations. Moreover, given two faces, $F$ and $G$, generated by different subsets of the same pure and perfectly distinguishable set of states, one has $P_FP_G=P_{F\cap G}$.

\section{Useful consequences of our principles}
\subsection{Uniqueness of distinguishing measurement} \label{Sym}
Strong symmetry (together with the no restriction hypothesis, which says that all mathematically well-defined measurements are physical) implies that, given any set of pure and perfectly distinguishable states $\{|i)\}$, there exists a unique measurement $\{(j|\}$ such that, \[(i|j)=\delta_{ij}.\] See \cite{muller2012structure,barnum2014higher} for details. Moreover, for every set $\{(e_j|\}$ such that $(e_j|i)=\alpha_j \delta_{ij}$, it holds that \[(e_j|=\alpha_j(j|.\]

\subsection{Purifications of completely mixed states are dynamically faithful}
As mentioned in section~\ref{framework}, purification implies that
there exist \emph{completely mixed} states. Purification implies that
there exists a state $|\psi)$ that purifies such a completely mixed
state:
\[\begin{tikzpicture}
	\begin{pgfonlayer}{nodelayer}
		\node [style=none] (0) at (1, -0.25) {};
		\node [style=none] (1) at (1, 1.25) {};
		\node [style=none] (2) at (1, 1.75) {};
		\node [style=none] (3) at (1, -0.75) {};
		\node [style=trace] (4) at (2.5, -0.25) {};
		\node [style=none] (5) at (2.5, 1.25) {};
		\node [style=none] (6) at (1, -0.75) {};
		\node [style=none] (7) at (0.5, 0.5) {$\psi$};
		\node [style=none] (8) at (3.75, 0.75) {$=$};
		\node [style=traceState] (9) at (5, 0.75) {};
		\node [style=none] (10) at (6.5, 0.75) {};
		\node [style=none] (11) at (7.25, -0) {};
	\end{pgfonlayer}
	\begin{pgfonlayer}{edgelayer}
		\draw [bend right=90, looseness=1.25] (2.center) to (3.center);
		\draw (2.center) to (3.center);
		\draw (1.center) to (5.center);
		\draw (0.center) to (4);
		\draw (9) to (10.center);
	\end{pgfonlayer}
\end{tikzpicture} \]
This is unique up to reversible transformation. We denote a particular choice of this purification as,
\[\begin{tikzpicture}
	\begin{pgfonlayer}{nodelayer}
		\node [style=none] (0) at (4.75, -0.25) {};
		\node [style=none] (1) at (4.75, 1.25) {};
		\node [style=none] (2) at (4.75, 1.75) {};
		\node [style=none] (3) at (4.75, -0.75) {};
		\node [style=none] (4) at (6.25, -0.25) {};
		\node [style=none] (5) at (6.25, 1.25) {};
		\node [style=none] (6) at (4.75, -0.75) {};
		\node [style=none] (7) at (4.25, 0.5) {$\psi$};
		\node [style=none] (8) at (2.5, 0.5) {$:=$};
		\node [style=none] (9) at (0.75, 1.25) {};
		\node [style=none] (10) at (1.25, -0.25) {};
		\node [style=none] (11) at (0.75, -0.25) {};
		\node [style=none] (12) at (1.25, 1.25) {};
		\node [style=none] (13) at (7, -0) {};
	\end{pgfonlayer}
	\begin{pgfonlayer}{edgelayer}
		\draw [bend right=90, looseness=1.25] (2.center) to (3.center);
		\draw (2.center) to (3.center);
		\draw (1.center) to (5.center);
		\draw (0.center) to (4.center);
		\draw (9.center) to (12.center);
		\draw (11.center) to (10.center);
		\draw [bend right=90, looseness=1.75] (9.center) to (11.center);
	\end{pgfonlayer}
\end{tikzpicture} \]
Purifications of completely mixed states are called \emph{dynamically faithful} states \cite{chiribella2010probabilistic} and, due to the constraints on parallel composition imposed in section~\ref{framework}, must satisfy the following important condition \cite{chiribella2010probabilistic}:
\[\begin{tikzpicture}
	\begin{pgfonlayer}{nodelayer}
		\node [style=none] (0) at (5.5, 7.25) {$=$};
		\node [style=none] (1) at (0.7499999, 7) {};
		\node [style=none] (2) at (1.25, 5.25) {};
		\node [style=none] (3) at (0.7499997, 5.25) {};
		\node [style=none] (4) at (1.25, 7) {};
		\node [style=none] (5) at (1.25, 9) {};
		\node [style=none] (6) at (1.25, 6) {};
		\node [style=none] (7) at (2.75, 6) {};
		\node [style=none] (8) at (2.75, 9) {};
		\node [style=none] (9) at (2, 7.5) {$T$};
		\node [style=none] (10) at (1.25, 8.5) {};
		\node [style=none] (11) at (1.25, 7.5) {};
		\node [style=none] (12) at (0, 8.5) {};
		\node [style=none] (13) at (3.75, 7) {};
		\node [style=none] (14) at (3.75, 8.5) {};
		\node [style=none] (15) at (2.75, 8.5) {};
		\node [style=none] (16) at (2.75, 7) {};
		\node [style=none] (17) at (3.75, 5.25) {};
		\node [style=none] (18) at (8.25, 6) {};
		\node [style=none] (19) at (8.25, 5.25) {};
		\node [style=none] (20) at (9.75, 7) {};
		\node [style=none] (21) at (8.25, 8.5) {};
		\node [style=none] (22) at (8.25, 9) {};
		\node [style=none] (23) at (9.75, 8.5) {};
		\node [style=none] (24) at (10.75, 5.25) {};
		\node [style=none] (25) at (9.75, 6) {};
		\node [style=none] (26) at (9.000001, 7.5) {$T'$};
		\node [style=none] (27) at (6.999999, 8.5) {};
		\node [style=none] (28) at (7.75, 7) {};
		\node [style=none] (29) at (8.25, 7.5) {};
		\node [style=none] (30) at (10.75, 8.5) {};
		\node [style=none] (31) at (9.75, 9) {};
		\node [style=none] (32) at (8.25, 7) {};
		\node [style=none] (33) at (7.75, 5.25) {};
		\node [style=none] (34) at (10.75, 7) {};
		\node [style=none] (35) at (2, 3.5) {$\implies$};
		\node [style=none] (36) at (3.75, -0.7499997) {};
		\node [style=none] (37) at (2.749999, -0) {};
		\node [style=none] (38) at (13.25, 1.75) {};
		\node [style=none] (39) at (11.5, 0.7499999) {$T'$};
		\node [style=none] (40) at (12.25, 0.25) {};
		\node [style=none] (41) at (5.25, 1.75) {};
		\node [style=none] (42) at (5.25, 2.25) {};
		\node [style=none] (43) at (12.25, -0.7499999) {};
		\node [style=none] (44) at (3.75, 1.75) {};
		\node [style=none] (45) at (3.75, -0) {};
		\node [style=none] (46) at (12.25, 2.25) {};
		\node [style=none] (47) at (9.500001, 1.75) {};
		\node [style=none] (48) at (3.75, 0.7499997) {};
		\node [style=none] (49) at (10.75, 1.75) {};
		\node [style=none] (50) at (5.25, 0.25) {};
		\node [style=none] (51) at (2.5, 1.75) {};
		\node [style=none] (52) at (3.75, 2.25) {};
		\node [style=none] (53) at (8, 0.5000001) {$=$};
		\node [style=none] (54) at (13.25, 0.2499996) {};
		\node [style=none] (55) at (10.75, -0) {};
		\node [style=none] (56) at (9.750001, -0) {};
		\node [style=none] (57) at (6.25, 1.75) {};
		\node [style=none] (58) at (4.5, 0.7499999) {$T$};
		\node [style=none] (59) at (12.25, 1.75) {};
		\node [style=none] (60) at (10.75, -0.7499997) {};
		\node [style=none] (61) at (6.249999, 0.2499996) {};
		\node [style=none] (62) at (5.25, -0.7499999) {};
		\node [style=none] (63) at (10.75, 0.7499997) {};
		\node [style=none] (64) at (10.75, 2.25) {};
		\node [style=none] (66) at (2.749999, -0) {};
		\node [style=none] (67) at (15.75, -0) {};
	\end{pgfonlayer}
	\begin{pgfonlayer}{edgelayer}
		\draw (1.center) to (4.center);
		\draw (3.center) to (2.center);
		\draw [bend right=90, looseness=1.75] (1.center) to (3.center);
		\draw (5.center) to (6.center);
		\draw (6.center) to (7.center);
		\draw (7.center) to (8.center);
		\draw (8.center) to (5.center);
		\draw (12.center) to (10.center);
		\draw (15.center) to (14.center);
		\draw (16.center) to (13.center);
		\draw (2.center) to (17.center);
		\draw (28.center) to (32.center);
		\draw (33.center) to (19.center);
		\draw [bend right=90, looseness=1.75] (28.center) to (33.center);
		\draw (22.center) to (18.center);
		\draw (18.center) to (25.center);
		\draw (25.center) to (31.center);
		\draw (31.center) to (22.center);
		\draw (27.center) to (21.center);
		\draw (23.center) to (30.center);
		\draw (20.center) to (34.center);
		\draw (19.center) to (24.center);
		\draw (37) to (45.center);
		\draw (52.center) to (36.center);
		\draw (36.center) to (62.center);
		\draw (62.center) to (42.center);
		\draw (42.center) to (52.center);
		\draw (51.center) to (44.center);
		\draw (41.center) to (57.center);
		\draw (50.center) to (61.center);
		\draw (64.center) to (60.center);
		\draw (60.center) to (43.center);
		\draw (43.center) to (46.center);
		\draw (46.center) to (64.center);
		\draw (47.center) to (49.center);
		\draw (56) to (55.center);
		\draw (59.center) to (38.center);
		\draw (40.center) to (54.center);
	\end{pgfonlayer}
\end{tikzpicture}  \]
As a special case, of course the purification of the \emph{maximally
  mixed} state is dynamically faithful.  In our applications, however,
any dynamically faithful state, purifying some completely mixed state,
will do.
\section{Proof of theorem~\ref{subroutine}} \label{Proof of subroutine}

\begin{proof}
It was shown in \cite{chiribella2010probabilistic} that the
purification principle implies the ability to dilate any
transformation to a reversible one. We use this fact in the
construction of the circuit $\{G_{|x|}\}$. Our construction is
equivalent to the one employed in the quantum case by
\cite{bennett1997strengths}.
 
In the construction, each $G_{|x|}$ is given by: 
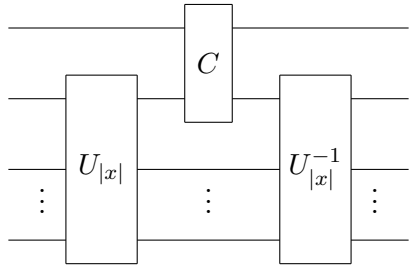
\[\begin{tikzpicture}[scale=1.25]
	\begin{pgfonlayer}{nodelayer}
		\node [style=none] (0) at (-0.5, 4) {};
		\node [style=none] (1) at (-0.5, 2.5) {};
		\node [style=none] (2) at (-0.5, 1) {};
		\node [style=none] (3) at (-0.5, -0.5) {};
		\node [style=none] (4) at (0.25, 0.5) {$\vdots$};
		\node [style=none] (5) at (0.75, 3) {};
		\node [style=none] (6) at (0.75, -1) {};
		\node [style=none] (7) at (2.25, -1) {};
		\node [style=none] (8) at (2.25, 3) {};
		\node [style=none] (9) at (1.5, 1) {$U_{|x|}$};
		\node [style=none] (10) at (0.75, 2.5) {};
		\node [style=none] (11) at (0.75, 1) {};
		\node [style=none] (12) at (0.75, -0.5) {};
		\node [style=none] (13) at (2.25, -0.5) {};
		\node [style=none] (14) at (2.25, 1) {};
		\node [style=none] (15) at (2.25, 2.5) {};
		\node [style=none] (16) at (3.25, 2.5) {};
		\node [style=none] (17) at (3.25, 2) {};
		\node [style=none] (18) at (4.25, 2) {};
		\node [style=none] (19) at (3.25, 4) {};
		\node [style=none] (20) at (4.25, 4) {};
		\node [style=none] (21) at (4.25, 2.5) {};
		\node [style=none] (22) at (4.25, 4.5) {};
		\node [style=none] (23) at (3.25, 4.5) {};
		\node [style=none] (24) at (3.75, 3.25) {$C$};
		\node [style=none] (25) at (5.25, -0.5) {};
		\node [style=none] (26) at (6.75, 1) {};
		\node [style=none] (27) at (6, 1) {$U^{-1}_{|x|}$};
		\node [style=none] (28) at (3.75, 0.5) {$\vdots$};
		\node [style=none] (29) at (5.25, -1) {};
		\node [style=none] (30) at (6.75, 3) {};
		\node [style=none] (31) at (6.75, 2.5) {};
		\node [style=none] (32) at (7.25, 0.5) {$\vdots$};
		\node [style=none] (33) at (6.75, -0.5) {};
		\node [style=none] (34) at (5.25, 1) {};
		\node [style=none] (35) at (5.25, 2.5) {};
		\node [style=none] (36) at (5.25, 3) {};
		\node [style=none] (37) at (6.75, -1) {};
		\node [style=none] (38) at (8, 1) {};
	    \node [style=none] (39) at (8, 2.5) {};
		\node [style=none] (40) at (8, -0.5) {};
		\node [style=none] (41) at (8, 4) {};
	\end{pgfonlayer}
	\begin{pgfonlayer}{edgelayer}
		\draw (5.center) to (8.center);
		\draw (8.center) to (7.center);
		\draw (7.center) to (6.center);
		\draw (6.center) to (5.center);
		\draw (36.center) to (30.center);
		\draw (30.center) to (37.center);
		\draw (37.center) to (29.center);
		\draw (29.center) to (36.center);
		\draw (0) to (19.center);
		\draw (23.center) to (17.center);
		\draw (17.center) to (18.center);
		\draw (18.center) to (22.center);
		\draw (22.center) to (23.center);
		\draw (15.center) to (16.center);
		\draw (10.center) to (1);
		\draw (2) to (11.center);
		\draw (3) to (12.center);
		\draw (13.center) to (25.center);
		\draw (14.center) to (34.center);
		\draw (21.center) to (35.center);
		\draw (20.center) to (41);
		\draw (31.center) to (39);
		\draw (26.center) to (38);
		\draw (33.center) to (40);
	\end{pgfonlayer}
\end{tikzpicture}\]
where $U_{|x|}$ is the reversible transformation which
dilates\footnote{Recall that that the circuit family
  $\{U_{|x|}\}$, with $U_{|x|}$ a reversible transformation which
  dilates $A_{|x|}$ for each ${|x|}$, consists of poly-size uniform
  circuits.} the \textbf{BGP} algorithm $A_{|x|}$
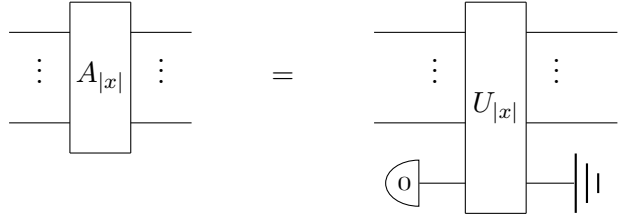
\[\begin{tikzpicture}[scale=1.6]
	\begin{pgfonlayer}{nodelayer}
		\node [style=none] (0) at (0, 2) {};
		\node [style=none] (1) at (1, 2) {};
		\node [style=none] (2) at (0, 0.5) {};
		\node [style=none] (3) at (1, 0.5) {};
		\node [style=none] (4) at (1, 2.5) {};
		\node [style=none] (5) at (0.5, 1.5) {$\vdots$};
		\node [style=none] (6) at (1, -0) {};
		\node [style=none] (7) at (2, -0) {};
		\node [style=none] (8) at (2, 2.5) {};
		\node [style=none] (9) at (2, 2) {};
		\node [style=none] (10) at (3, 2) {};
		\node [style=none] (11) at (2, 0.5) {};
		\node [style=none] (12) at (3, 0.5) {};
		\node [style=none] (13) at (2.5, 1.5) {$\vdots$};
		\node [style=none] (14) at (1.5, 1.25) {$A_{|x|}$};
		\node [style=none] (15) at (4.5, 1.25) {$=$};
		\node [style=none] (16) at (8.5, 2) {};
		\node [style=none] (17) at (7.5, 0.5) {};
		\node [style=none] (18) at (7.5, 2.5) {};
		\node [style=none] (19) at (8.5, -1) {};
		\node [style=none] (20) at (9, 1.5) {$\vdots$};
		\node [style=none] (21) at (10, 2) {};
		\node [style=none] (22) at (10, 0.5) {};
		\node [style=none] (23) at (6, 2) {};
		\node [style=none] (24) at (7.5, 2) {};
		\node [style=none] (25) at (7.5, -1) {};
		\node [style=none] (26) at (6, 0.5) {};
		\node [style=none] (27) at (8, 0.75) {$U_{|x|}$};
		\node [style=none] (28) at (8.5, 2.5) {};
		\node [style=none] (29) at (7, 1.5) {$\vdots$};
		\node [style=none] (30) at (8.5, 0.5) {};
		\node [style=none] (31) at (7.5, -0.5) {};
		\node [style=cpoint] (32) at (6.5, -0.5) {$0$};
		\node [style=none] (33) at (8.5, -0.5) {};
		\node [style=trace] (34) at (9.5, -0.5) {};
	\end{pgfonlayer}
	\begin{pgfonlayer}{edgelayer}
		\draw (0.center) to (1.center);
		\draw (4.center) to (6.center);
		\draw (6.center) to (7.center);
		\draw (7.center) to (8.center);
		\draw (8.center) to (4.center);
		\draw (2.center) to (3.center);
		\draw (11.center) to (12.center);
		\draw (9.center) to (10.center);
		\draw (23.center) to (24.center);
		\draw (18.center) to (25.center);
		\draw (25.center) to (19.center);
		\draw (19.center) to (28.center);
		\draw (28.center) to (18.center);
		\draw (26.center) to (17.center);
		\draw (30.center) to (22.center);
		\draw (16.center) to (21.center);
		\draw (32) to (31.center);
		\draw (33.center) to (34);
	\end{pgfonlayer}
\end{tikzpicture}\]
and $C$ is ``controlled bit-flip'', or ``generalised CNOT'': 
a reversible controlled transformation with the lower
system as the control
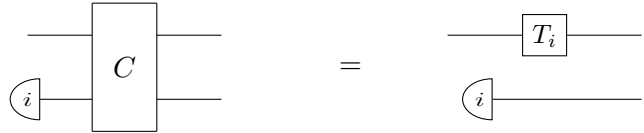
\[
\begin{tikzpicture}[scale=1.7]
	\begin{pgfonlayer}{nodelayer}
		\node [style=none] (0) at (0.9999999, -0) {};
		\node [style=none] (1) at (2, -0) {};
		\node [style=none] (2) at (0.9999999, 1.5) {};
		\node [style=none] (3) at (2, 1.5) {};
		\node [style=none] (4) at (2, 2) {};
		\node [style=none] (5) at (0.9999999, 2) {};
		\node [style=none] (6) at (1.5, 0.9999999) {$C$};
		\node [style=none] (7) at (5, 0.9999999) {$=$};
		\node [style=none] (8) at (0, 1.5) {};
		\node [style=none] (9) at (3, 1.5) {};
		\node [style=none] (10) at (0.9999999, 0.4999999) {};
		\node [style=none] (11) at (2, 0.4999999) {};
		\node [style=none] (12) at (3, 0.4999999) {};
		\node [style=cpoint] (13) at (0, 0.4999999) {$i$};
		\node [style=none] (14) at (6.499999, 1.5) {};
		\node [style=none] (15) at (9.500001, 0.4999999) {};
		\node [style=cpoint] (16) at (6.999999, 0.4999999) {$i$};
		\node [style=none] (17) at (9.500001, 1.5) {};
		\node [style={small box}] (18) at (7.999999, 1.5) {$T_i$};
	\end{pgfonlayer}
	\begin{pgfonlayer}{edgelayer}
		\draw (5.center) to (0.center);
		\draw (0.center) to (1.center);
		\draw (1.center) to (4.center);
		\draw (4.center) to (5.center);
		\draw (8.center) to (2.center);
		\draw (3.center) to (9.center);
		\draw (11.center) to (12.center);
		\draw (13) to (10.center);
		\draw (16) to (15.center);
		\draw (14.center) to (18);
		\draw (18) to (17.center);
	\end{pgfonlayer}
\end{tikzpicture}
\]
with $|i)\in\{|0), |1)\}$, $T_0=\mathds{I}$, and where $T_1$ acts as
$T_1|i)=|i\oplus 1)$.

To show that the family $G_{|x|}$ functions as an oracle with high
probability, thereby proving theorem~\ref{subroutine}, we will show
that the probability corresponding to the following closed circuit
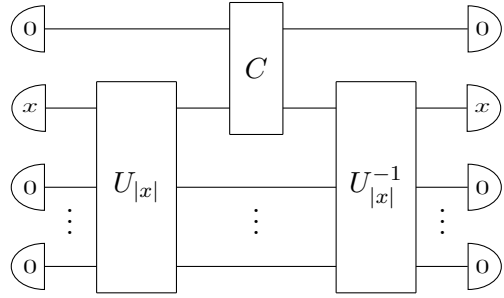
\[\begin{tikzpicture}[scale=1.4]
	\begin{pgfonlayer}{nodelayer}
		\node [style=cpoint] (0) at (-0.5, 4) {$0$};
		\node [style=cpoint] (1) at (-0.5, 2.5) {$x$};
		\node [style=cpoint] (2) at (-0.5, 1) {$0$};
		\node [style=cpoint] (3) at (-0.5, -0.5) {$0$};
		\node [style=none] (4) at (0.25, 0.5) {$\vdots$};
		\node [style=none] (5) at (0.75, 3) {};
		\node [style=none] (6) at (0.75, -1) {};
		\node [style=none] (7) at (2.25, -1) {};
		\node [style=none] (8) at (2.25, 3) {};
		\node [style=none] (9) at (1.5, 1) {$U_{|x|}$};
		\node [style=none] (10) at (0.75, 2.5) {};
		\node [style=none] (11) at (0.75, 1) {};
		\node [style=none] (12) at (0.75, -0.5) {};
		\node [style=none] (13) at (2.25, -0.5) {};
		\node [style=none] (14) at (2.25, 1) {};
		\node [style=none] (15) at (2.25, 2.5) {};
		\node [style=none] (16) at (3.25, 2.5) {};
		\node [style=none] (17) at (3.25, 2) {};
		\node [style=none] (18) at (4.25, 2) {};
		\node [style=none] (19) at (3.25, 4) {};
		\node [style=none] (20) at (4.25, 4) {};
		\node [style=none] (21) at (4.25, 2.5) {};
		\node [style=none] (22) at (4.25, 4.5) {};
		\node [style=none] (23) at (3.25, 4.5) {};
		\node [style=none] (24) at (3.75, 3.25) {$C$};
		\node [style=none] (25) at (5.25, -0.5) {};
		\node [style=none] (26) at (6.75, 1) {};
		\node [style=none] (27) at (6, 1) {$U^{-1}_{|x|}$};
		\node [style=none] (28) at (3.75, 0.5) {$\vdots$};
		\node [style=none] (29) at (5.25, -1) {};
		\node [style=none] (30) at (6.75, 3) {};
		\node [style=none] (31) at (6.75, 2.5) {};
		\node [style=none] (32) at (7.25, 0.5) {$\vdots$};
		\node [style=none] (33) at (6.75, -0.5) {};
		\node [style=none] (34) at (5.25, 1) {};
		\node [style=none] (35) at (5.25, 2.5) {};
		\node [style=none] (36) at (5.25, 3) {};
		\node [style=none] (37) at (6.75, -1) {};
		\node [style=cocpoint] (38) at (8, 1) {$0$};
		\node [style=cocpoint] (39) at (8, 2.5) {$x$};
		\node [style=cocpoint] (40) at (8, -0.5) {$0$};
		\node [style=cocpoint] (41) at (8, 4) {$0$};
	\end{pgfonlayer}
	\begin{pgfonlayer}{edgelayer}
		\draw (5.center) to (8.center);
		\draw (8.center) to (7.center);
		\draw (7.center) to (6.center);
		\draw (6.center) to (5.center);
		\draw (36.center) to (30.center);
		\draw (30.center) to (37.center);
		\draw (37.center) to (29.center);
		\draw (29.center) to (36.center);
		\draw (0) to (19.center);
		\draw (23.center) to (17.center);
		\draw (17.center) to (18.center);
		\draw (18.center) to (22.center);
		\draw (22.center) to (23.center);
		\draw (15.center) to (16.center);
		\draw (10.center) to (1);
		\draw (2) to (11.center);
		\draw (3) to (12.center);
		\draw (13.center) to (25.center);
		\draw (14.center) to (34.center);
		\draw (21.center) to (35.center);
		\draw (20.center) to (41);
		\draw (31.center) to (39);
		\draw (26.center) to (38);
		\draw (33.center) to (40);
	\end{pgfonlayer}
\end{tikzpicture}\]
is greater than or equal to $1-2^{-q(|x|)}$, for some polynomial
$q(|x|)$, when the algorithm $A_{|x|}$ accepts\footnote{That is, when
  $x$ is in the language decided by the algorithm.} the input $x$.


We choose the dynamically faithful state to satisfy
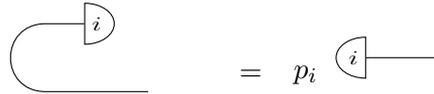
\[\begin{tikzpicture}[scale=1.8]
	\begin{pgfonlayer}{nodelayer}
		\node [style=none] (0) at (0.4999999, 0.9999999) {};
		\node [style=cocpoint] (1) at (1.250001, 0.9999999) {$i$};
		\node [style=none] (2) at (3.5, 0.25) {$=$};
		\node [style=none] (7) at (4.3, 0.25) {$p_i$};
		\node [style=cpoint] (3) at (5, 0.4999999) {$i$};
		\node [style=none] (4) at (6.25, 0.4999999) {};
		\node [style=none] (5) at (2, -0) {};
		\node [style=none] (6) at (0.4999999, -0) {};
	\end{pgfonlayer}
	\begin{pgfonlayer}{edgelayer}
		\draw (0.center) to (1);
		\draw (3) to (4.center);
		\draw [bend right=90, looseness=1.75] (0.center) to (6.center);
		\draw (6.center) to (5.center);
	\end{pgfonlayer}
\end{tikzpicture}\]
 where $p_i\in (0,1]$ and $\sum_i p_i=1$, which can always be achieved without loss of generality (see theorem $6$ and corollary $9$ from \cite{chiribella2010probabilistic}).

We first show that $C$ satisfies
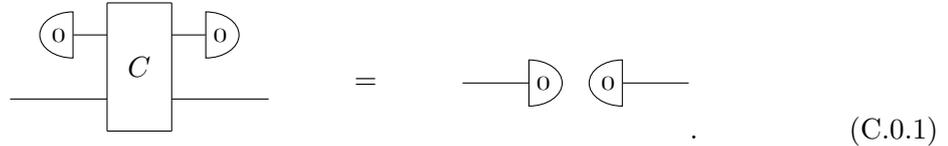
\begin{equation}\label{equation1}\begin{tikzpicture}[scale=1.7]
	\begin{pgfonlayer}{nodelayer}
		\node [style=cpoint] (0) at (0.750001, 1.5) {$0$};
		\node [style=none] (1) at (1.5, 0.4999999) {};
		\node [style=none] (2) at (1.5, -0) {};
		\node [style=none] (3) at (2.5, -0) {};
		\node [style=none] (4) at (1.5, 1.5) {};
		\node [style=none] (5) at (2.5, 1.5) {};
		\node [style=none] (6) at (2.5, 0.4999999) {};
		\node [style=none] (7) at (2.5, 2) {};
		\node [style=none] (8) at (1.5, 2) {};
		\node [style=none] (9) at (2, 0.9999999) {$C$};
		\node [style=cocpoint] (10) at (3.250001, 1.5) {$0$};
		\node [style=none] (11) at (0, 0.4999999) {};
		\node [style=none] (12) at (4, 0.4999999) {};
		\node [style=none] (13) at (5.499999, 0.7499999) {$=$};
		\node [style=none] (14) at (6.999999, 0.7499999) {};
		\node [style=cocpoint] (15) at (8.25, 0.7499999) {$0$};
		\node [style=cpoint] (16) at (9.25, 0.7499999) {$0$};
		\node [style=none] (17) at (10.5, 0.7499999) {};
	\end{pgfonlayer}
	\begin{pgfonlayer}{edgelayer}
		\draw (0) to (4.center);
		\draw (8.center) to (2.center);
		\draw (2.center) to (3.center);
		\draw (3.center) to (7.center);
		\draw (7.center) to (8.center);
		\draw (5.center) to (10);
		\draw (11.center) to (1.center);
		\draw (6.center) to (12.center);
		\draw (14.center) to (15);
		\draw (16) to (17.center);
	\end{pgfonlayer}
\end{tikzpicture}.
\end{equation}
To see this note that uniqueness of measurement (both of the following
states give probability $p_0$ for $(0|(0|$, and probability zero for
each of $(0|(1|, (1|(0|, $ and $(1|(1|$) implies
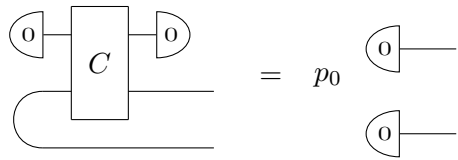
\[\begin{tikzpicture}[scale=1.5]
	\begin{pgfonlayer}{nodelayer}
		\node [style=cpoint] (0) at (0.2499996, 2) {$0$};
		\node [style=none] (1) at (1, 1) {};
		\node [style=none] (2) at (1, 2.5) {};
		\node [style=none] (3) at (2, 2.5) {};
		\node [style=none] (4) at (1, 2) {};
		\node [style=none] (5) at (2, 2) {};
		\node [style=none] (6) at (2, 1) {};
		\node [style=none] (7) at (2, 0.5000001) {};
		\node [style=none] (8) at (1, 0.5000001) {};
		\node [style=none] (9) at (1.5, 1.5) {$C$};
		\node [style=cocpoint] (10) at (2.749999, 2) {$0$};
		\node [style=none] (11) at (0.5000001, 1) {};
		\node [style=none] (12) at (3.5, 1) {};
		\node [style=none] (13) at (4.5, 1.25) {$=$};
			\node [style=none] (20) at (5.5, 1.25) {$p_0$};
		\node [style=cpoint] (14) at (6.5, 0.2500001) {$0$};
		\node [style=none] (15) at (7.75, 0.2500001) {};
		\node [style=cpoint] (16) at (6.5, 1.75) {$0$};
		\node [style=none] (17) at (7.75, 1.75) {};
		\node [style=none] (18) at (0.5000001, -0) {};
		\node [style=none] (19) at (3.5, -0) {};
	\end{pgfonlayer}
	\begin{pgfonlayer}{edgelayer}
		\draw (0) to (4.center);
		\draw (8.center) to (2.center);
		\draw (2.center) to (3.center);
		\draw (3.center) to (7.center);
		\draw (7.center) to (8.center);
		\draw (5.center) to (10);
		\draw (11.center) to (1.center);
		\draw (6.center) to (12.center);
		\draw (14) to (15.center);
		\draw (16) to (17.center);
		\draw (18.center) to (19.center);
		\draw [bend right=90, looseness=1.75] (11.center) to (18.center);
	\end{pgfonlayer}
\end{tikzpicture}\]
From our choice of dynamically faithful state, it then follows that
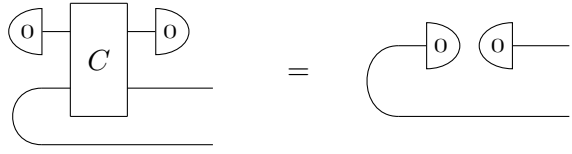
\[\begin{tikzpicture}[scale=1.5]
	\begin{pgfonlayer}{nodelayer}
		\node [style=cpoint] (0) at (0.2499996, 2) {$0$};
		\node [style=none] (1) at (1, 1) {};
		\node [style=none] (2) at (1, 2.5) {};
		\node [style=none] (3) at (2, 2.5) {};
		\node [style=none] (4) at (1, 2) {};
		\node [style=none] (5) at (2, 2) {};
		\node [style=none] (6) at (2, 1) {};
		\node [style=none] (7) at (2, 0.5000001) {};
		\node [style=none] (8) at (1, 0.5000001) {};
		\node [style=none] (9) at (1.5, 1.5) {$C$};
		\node [style=cocpoint] (10) at (2.749999, 2) {$0$};
		\node [style=none] (11) at (0.5000001, 1) {};
		\node [style=none] (12) at (3.5, 1) {};
		\node [style=none] (13) at (5, 1.25) {$=$};
		\node [style=none] (14) at (9.750001, 0.5000001) {};
		\node [style=cpoint] (15) at (8.499999, 1.75) {$0$};
		\node [style=none] (16) at (9.750001, 1.75) {};
		\node [style=none] (17) at (0.5000001, -0) {};
		\node [style=none] (18) at (3.5, -0) {};
		\node [style=cocpoint] (19) at (7.500001, 1.75) {$0$};
		\node [style=none] (20) at (6.75, 1.75) {};
		\node [style=none] (21) at (6.75, 0.5000001) {};
	\end{pgfonlayer}
	\begin{pgfonlayer}{edgelayer}
		\draw (0) to (4.center);
		\draw (8.center) to (2.center);
		\draw (2.center) to (3.center);
		\draw (3.center) to (7.center);
		\draw (7.center) to (8.center);
		\draw (5.center) to (10);
		\draw (11.center) to (1.center);
		\draw (6.center) to (12.center);
		\draw (15) to (16.center);
		\draw (17.center) to (18.center);
		\draw [bend right=90, looseness=1.75] (11.center) to (17.center);
		\draw (20.center) to (19);
		\draw [bend right=90, looseness=1.50] (20.center) to (21.center);
		\draw (21.center) to (14.center);
	\end{pgfonlayer}
\end{tikzpicture} \]
Dynamical faithfulness then gives equation  (\ref{equation1}). 

Secondly, we write 
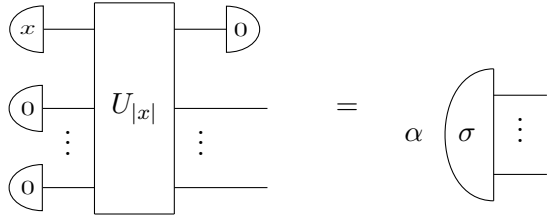
\[\begin{tikzpicture}[scale=1.4]
	\begin{pgfonlayer}{nodelayer}
		\node [style=cpoint] (0) at (-0.5, 2.5) {$x$};
		\node [style=cpoint] (1) at (-0.5, 1) {$0$};
		\node [style=cpoint] (2) at (-0.5, -0.5) {$0$};
		\node [style=none] (3) at (0.25, 0.5) {$\vdots$};
		\node [style=none] (4) at (0.75, 3) {};
		\node [style=none] (5) at (0.75, -1) {};
		\node [style=none] (6) at (2.25, -1) {};
		\node [style=none] (7) at (2.25, 3) {};
		\node [style=none] (8) at (1.5, 1) {$U_{|x|}$};
		\node [style=none] (9) at (0.75, 2.5) {};
		\node [style=none] (10) at (0.75, 1) {};
		\node [style=none] (11) at (0.75, -0.5) {};
		\node [style=none] (12) at (2.25, -0.5) {};
		\node [style=none] (13) at (2.25, 1) {};
		\node [style=none] (14) at (2.25, 2.5) {};
		\node [style=none] (15) at (4, -0.5) {};
		\node [style=none] (16) at (2.75, 0.5) {$\vdots$};
		\node [style=none] (17) at (4, 1) {};
		\node [style=cocpoint] (18) at (3.5, 2.5) {$0$};
		\node [style=none] (19) at (5.5, 1) {$=$};
		\node [style=none] (20) at (8.25, 1.75) {};
		\node [style=none] (21) at (8.25, 1.25) {};
		\node [style=none] (22) at (8.25, -0.25) {};
		\node [style=none] (23) at (8.25, -0.75) {};
		\node [style=none] (24) at (7.75, 0.5) {$\sigma$};
		\node [style=none] (25) at (6.75, 0.5) {$\alpha$};
		\node [style=none] (26) at (9.25, 1.25) {};
		\node [style=none] (27) at (9.25, -0.25) {};
		\node [style=none] (28) at (8.75, 0.75) {$\vdots$};
	\end{pgfonlayer}
	\begin{pgfonlayer}{edgelayer}
		\draw (4.center) to (7.center);
		\draw (7.center) to (6.center);
		\draw (6.center) to (5.center);
		\draw (5.center) to (4.center);
		\draw (9.center) to (0);
		\draw (1) to (10.center);
		\draw (2) to (11.center);
		\draw (12.center) to (15.center);
		\draw (13.center) to (17.center);
		\draw (14.center) to (18);
		\draw [bend right=90, looseness=1.25] (20.center) to (23.center);
		\draw (20.center) to (23.center);
		\draw (21.center) to (26.center);
		\draw (22.center) to (27.center);
	\end{pgfonlayer}
\end{tikzpicture}\]
where $|\sigma)$ is a normalised state and $\alpha\in (0,1]$. Our choice of acceptance condition, together with the fact that $U$ is a dilation of the algorithm $A$, results in
\begin{equation}
\label{equation2}
\begin{tikzpicture}[scale=1.4]
	\begin{pgfonlayer}{nodelayer}
		\node [style=cpoint] (0) at (-0.5, 2.5) {$x$};
		\node [style=cpoint] (1) at (-0.5, 1) {$0$};
		\node [style=cpoint] (2) at (-0.5, -0.5) {$0$};
		\node [style=none] (3) at (0.25, 0.5) {$\vdots$};
		\node [style=none] (4) at (0.75, 3) {};
		\node [style=none] (5) at (0.75, -1) {};
		\node [style=none] (6) at (2.25, -1) {};
		\node [style=none] (7) at (2.25, 3) {};
		\node [style=none] (8) at (1.5, 1) {$U_{|x|}$};
		\node [style=none] (9) at (0.75, 2.5) {};
		\node [style=none] (10) at (0.75, 1) {};
		\node [style=none] (11) at (0.75, -0.5) {};
		\node [style=none] (12) at (2.25, -0.5) {};
		\node [style=none] (13) at (2.25, 1) {};
		\node [style=none] (14) at (2.25, 2.5) {};
		\node [style=none] (15) at (2.75, 0.5) {$\vdots$};
		\node [style=cocpoint] (16) at (3.5, 2.5) {$0$};
		\node [style=none] (17) at (5.5, 1) {$=$};
		\node [style=trace] (18) at (3.5, 1) {};
		\node [style=trace] (19) at (3.5, -0.5) {};
		\node [style=none] (20) at (6.5, 1) {$\alpha$};
		\node [style=none] (21) at (7.5, 1) {$=$};
		\node [style=none] (22) at (9.25, 1) {$P_x(acc)$};
	\end{pgfonlayer}
	\begin{pgfonlayer}{edgelayer}
		\draw (4.center) to (7.center);
		\draw (7.center) to (6.center);
		\draw (6.center) to (5.center);
		\draw (5.center) to (4.center);
		\draw (9.center) to (0);
		\draw (1) to (10.center);
		\draw (2) to (11.center);
		\draw (14.center) to (16);
		\draw (13.center) to (18);
		\draw (12.center) to (19);
	\end{pgfonlayer}
\end{tikzpicture}
\end{equation}
Combining  (\ref{equation1}) and (\ref{equation2}) gives
\[\begin{tikzpicture}[scale=1.2]
	\begin{pgfonlayer}{nodelayer}
		\node [style=cpoint] (0) at (-0.5, 4) {$0$};
		\node [style=cpoint] (1) at (-0.5, 2.5) {$x$};
		\node [style=cpoint] (2) at (-0.5, 1) {$0$};
		\node [style=cpoint] (3) at (-0.5, -0.5) {$0$};
		\node [style=none] (4) at (0.25, 0.5) {$\vdots$};
		\node [style=none] (5) at (0.75, 3) {};
		\node [style=none] (6) at (0.75, -1) {};
		\node [style=none] (7) at (2.25, -1) {};
		\node [style=none] (8) at (2.25, 3) {};
		\node [style=none] (9) at (1.5, 1) {$U_{|x|}$};
		\node [style=none] (10) at (0.75, 2.5) {};
		\node [style=none] (11) at (0.75, 1) {};
		\node [style=none] (12) at (0.75, -0.5) {};
		\node [style=none] (13) at (2.25, -0.5) {};
		\node [style=none] (14) at (2.25, 1) {};
		\node [style=none] (15) at (2.25, 2.5) {};
		\node [style=none] (16) at (3.25, 2.5) {};
		\node [style=none] (17) at (3.25, 2) {};
		\node [style=none] (18) at (4.25, 2) {};
		\node [style=none] (19) at (3.25, 4) {};
		\node [style=none] (20) at (4.25, 4) {};
		\node [style=none] (21) at (4.25, 2.5) {};
		\node [style=none] (22) at (4.25, 4.5) {};
		\node [style=none] (23) at (3.25, 4.5) {};
		\node [style=none] (24) at (3.75, 3.25) {$C$};
		\node [style=none] (25) at (5.25, -0.5) {};
		\node [style=none] (26) at (6.75, 1) {};
		\node [style=none] (27) at (6, 1) {$U^{-1}_{|x|}$};
		\node [style=none] (28) at (3.75, 0.5) {$\vdots$};
		\node [style=none] (29) at (5.25, -1) {};
		\node [style=none] (30) at (6.75, 3) {};
		\node [style=none] (31) at (6.75, 2.5) {};
		\node [style=none] (32) at (7.25, 0.5) {$\vdots$};
		\node [style=none] (33) at (6.75, -0.5) {};
		\node [style=none] (34) at (5.25, 1) {};
		\node [style=none] (35) at (5.25, 2.5) {};
		\node [style=none] (36) at (5.25, 3) {};
		\node [style=none] (37) at (6.75, -1) {};
		\node [style=cocpoint] (38) at (8, 1) {$0$};
		\node [style=cocpoint] (39) at (8, 2.5) {$x$};
		\node [style=cocpoint] (40) at (8, -0.5) {$0$};
		\node [style=cocpoint] (41) at (8, 4) {$0$};
		\node [style=none] (42) at (10, 2) {$=$};
		\node [style=none] (43) at (17.25, -0.2500001) {};
		\node [style=none] (44) at (15.75, 2.75) {};
		\node [style=none] (45) at (17.25, 1.25) {};
		\node [style=none] (46) at (17.25, 2.75) {};
		\node [style=none] (47) at (15.25, 0.7500001) {$\vdots$};
		\node [style=none] (48) at (17.75, 0.7500001) {$\vdots$};
		\node [style=none] (49) at (15.75, 3.25) {};
		\node [style=none] (50) at (15.75, -0.7500001) {};
		\node [style=cocpoint] (51) at (18.5, 2.75) {$x$};
		\node [style=none] (52) at (15.75, 1.25) {};
		\node [style=none] (53) at (17.25, 3.25) {};
		\node [style=cocpoint] (54) at (18.5, -0.2500001) {$0$};
		\node [style=none] (55) at (15.75, -0.2500001) {};
		\node [style=none] (56) at (16.5, 1.25) {$U^{-1}_{|x|}$};
		\node [style=none] (57) at (17.25, -0.7500001) {};
		\node [style=cocpoint] (58) at (18.5, 1.25) {$0$};
		\node [style=cpoint] (59) at (14.5, 2.75) {$0$};
		\node [style=none] (60) at (14.5, 1.25) {};
		\node [style=none] (61) at (14.5, -0.2500001) {};
		\node [style=none] (62) at (14.5, 1.75) {};
		\node [style=none] (63) at (14.5, -0.7500001) {};
		\node [style=none] (64) at (14, 0.4999999) {$\sigma$};
		\node [style=none] (65) at (12, 2) {$P_x(acc)$};
		\node [style=none] (66) at (15.75, -4.25) {$\vdots$};
		\node [style=none] (67) at (14.5, -4.5) {$\sigma$};
		\node [style=none] (68) at (15, -5.75) {};
		\node [style=none] (69) at (15, -3.75) {};
		\node [style=none] (70) at (12, -3.75) {$P_x(acc)^2$};
		\node [style=none] (71) at (10, -3.75) {$=$};
		\node [style=none] (72) at (16.25, -3.75) {};
		\node [style=none] (73) at (15, -3.25) {};
		\node [style=none] (74) at (16.25, -5.25) {};
		\node [style=none] (75) at (15, -5.25) {};
		\node [style=none] (76) at (16.25, -5.75) {};
		\node [style=none] (77) at (16.25, -5.25) {};
		\node [style=none] (78) at (16.25, -3.75) {};
		\node [style=none] (79) at (16.25, -3.25) {};
		\node [style=none] (80) at (16.75, -4.5) {$\sigma$};
	\end{pgfonlayer}
	\begin{pgfonlayer}{edgelayer}
		\draw (5.center) to (8.center);
		\draw (8.center) to (7.center);
		\draw (7.center) to (6.center);
		\draw (6.center) to (5.center);
		\draw (36.center) to (30.center);
		\draw (30.center) to (37.center);
		\draw (37.center) to (29.center);
		\draw (29.center) to (36.center);
		\draw (0) to (19.center);
		\draw (23.center) to (17.center);
		\draw (17.center) to (18.center);
		\draw (18.center) to (22.center);
		\draw (22.center) to (23.center);
		\draw (15.center) to (16.center);
		\draw (10.center) to (1);
		\draw (2) to (11.center);
		\draw (3) to (12.center);
		\draw (13.center) to (25.center);
		\draw (14.center) to (34.center);
		\draw (21.center) to (35.center);
		\draw (20.center) to (41);
		\draw (31.center) to (39);
		\draw (26.center) to (38);
		\draw (33.center) to (40);
		\draw (49.center) to (53.center);
		\draw (53.center) to (57.center);
		\draw (57.center) to (50.center);
		\draw (50.center) to (49.center);
		\draw (46.center) to (51);
		\draw (45.center) to (58);
		\draw (43.center) to (54);
		\draw (59) to (44.center);
		\draw [bend right=90, looseness=1.25] (62.center) to (63.center);
		\draw (60.center) to (52.center);
		\draw (61.center) to (55.center);
		\draw (62.center) to (63.center);
		\draw [bend right=90, looseness=1.25] (73.center) to (68.center);
		\draw (69.center) to (72.center);
		\draw (75.center) to (74.center);
		\draw (73.center) to (68.center);
		\draw [bend left=90, looseness=1.25] (79.center) to (76.center);
		\draw (79.center) to (76.center);
	\end{pgfonlayer}
\end{tikzpicture}\]
where the last line follows from self-duality.  By amplifying the
acceptance probability of the original algorithm $A$ (see
\cite{lee2015computation} for an in depth discussion of bounded error
efficient computation and amplifying acceptance probabilities), we can
ensure that when $x$ is in the language we have $P_x(acc)\geq
1-2^{-p(|x|)}$ for an arbitrary polynomial $p(|x|)$. Hence it follows
that $P_x(acc)^2\geq 1-2^{-p(|x|)+1}$. If $(\sigma |\sigma)=1$,
choosing $p(|x|)=q(|x|)+1$ completes the proof.

The case $(\sigma |\sigma) < 1$ can be easily dealt with. As
$|\sigma)$ and $(\sigma|$ can be efficiently prepared by a poly-size
circuit, the factor $(\sigma | \sigma)$ can be approximated by a
rational number to high accuracy (this is a consequence of the
computational uniformity condition required to define computation in
arbitrary physical theories, including quantum theory. See
\cite{lee2015computation} and \cite{lee2015proofs} for an expanded
discussion of this point). Hence one can write $(\sigma
|\sigma)=1-c2^{-w(|x|)}$, for $w$ a polynomial in the size of the
circuit and $c$ a constant natural number. One can always find a
polynomial $q$ such that
$\left(1-c2^{-w(|x|)}\right)\left(1-2^{-p(|x|)+1}\right) \geq
1-2^{-q(|x|)}$. This completes the proof.
\end{proof}

\end{appendices}

\bibliographystyle{ieeetr}
\bibliography{library_hbedit}

\end{document}